\documentclass[11pt]{article}
\usepackage[DIV13]{typearea}
\usepackage{amsmath}
\usepackage{amsfonts}
\usepackage{amssymb}
\usepackage[titletoc]{appendix}
\usepackage{array}
\usepackage{bbold} 
\usepackage{bbm}
\usepackage{slashed}
\usepackage{booktabs}
 \usepackage{cancel}
 \usepackage{subcaption}
 \usepackage{cite} 
\usepackage[usenames,dvipsnames]{color}
\usepackage{epsfig}
\usepackage{fancyhdr}
\usepackage[T1]{fontenc}
\usepackage{framed}
\usepackage[left=.9in, right=.9in]{geometry}
\usepackage{graphicx}
\usepackage{hhline} 
\RequirePackage[colorlinks=true,urlcolor=blue,anchorcolor=blue,citecolor=blue,filecolor=blue,
               linkcolor=blue,menucolor=blue,linktocpage=true,pdfproducer=medialab,pagebackref=true]{hyperref}
\usepackage[utf8]{inputenc}
\usepackage{mathrsfs}
\usepackage{mathtools}
\usepackage{dsfont}
\usepackage{multicol}
\usepackage{multirow}
\usepackage{indentfirst}
\usepackage{longtable}
 \usepackage{lscape}
\usepackage{pbox}
\usepackage{pdfpages}
\usepackage{rotating}
\usepackage{slashed}
\usepackage[DIV13]{typearea}
\usepackage[normalem]{ulem}
\usepackage{units}
\usepackage{enumitem}
\usepackage{soul}
\usepackage{placeins}
\usepackage{tikz}

\hypersetup{
  colorlinks   = true, 
  urlcolor     = magenta, 
  linkcolor    = blue, 
  citecolor   = blue 
}

\numberwithin{equation}{section}

\usepackage{csquotes}

\DeclareRobustCommand{\fourth}{\ensuremath{\frac{1}{4}}}

\DeclareRobustCommand{\alphaem}{\ensuremath{\alpha_{\mathrm{em}}}}

\newcommand{\blue}[1]{\color{blue}#1\color{black}}

\newcommand{\beq}{\begin{equation}}%
\newcommand{\eeq}{\end{equation}}%
\newcommand{\bea}{\begin{align}}%
\newcommand{\ena}{\end{align}}%

\usepackage{catchfile}
\newcommand{\getenv}[2][]{%
  \CatchFileEdef{\temp}{"|kpsewhich --var-value #2"}{}%
  \if\relax\detokenize{#1}\relax\temp\else\let#1\temp\fi
}
\getenv[\USER]{USER}

\usepackage{catchfile}

\allowdisplaybreaks
\newcommand{\m}{\mu}

\newcommand{\s}{\sigma}

\renewcommand{\m}{\mu}
\newcommand{\n}{\nu}

\renewcommand{\L}{\mathcal{L}}
\renewcommand{\O}{\mathbf{O}}

\renewcommand{\to}{\rightarrow}

\newcommand{\hc}{\mathrm{h.c.}}

\renewcommand{\[}{\begin{equation}}
\renewcommand{\]}{\end{equation}}
\definecolor{orange}{rgb}{1,0.5,0}

\newcommand{\LL}{\mathscr{L}}

\newcounter{diagram}

\newcommand{\email}[1]{\href{mailto:#1}{\tt #1}}

\renewcommand{\to}{\rightarrow}
\newcommand{\de}{\partial}
\newcommand{\nn}{\nonumber}
\renewcommand{\to}{\longrightarrow}

\renewcommand{\[}{\begin{equation}}
\renewcommand{\]}{\end{equation}}
\newcommand{\bmat}{\begin{pmatrix}}
\newcommand{\emat}{\end{pmatrix}}
\renewcommand{\to}{\longrightarrow}

\newcommand{\dmu}{\partial_\mu}
\newcommand{\dmup}{\partial^\mu}

\newcommand{\fpq}{f_{\text{PQ}}}

\newcommand{\BBd}{B_{\mu\nu}}

\newcommand{\WWd}{W_{\mu\nu}}

\def\cY{{\bf Y}}

\def\cY{{\bf Y}}

\definecolor{green}{rgb}{0.13, 0.55, 0.13}

\newcolumntype{C}[1]{>{\centering\let\newline\\\arraybackslash\hspace{0pt}}m{#1}}

\begin{document}
\vspace*{-1cm}
\phantom{hep-ph/***} 
{\flushright
{\blue{IFT-UAM/CSIC-18-110}}\hfill
{\blue{FTUAM-18-25}}
}
\vskip 1.5cm
\begin{center}
{\LARGE  Axion couplings to electroweak gauge bosons \\[0.5cm] }
\vskip .3cm
\end{center}
\vskip 0.5  cm
\begin{center}
{\large G.~Alonso\,-{\'A\!}lvarez}~$^{a)}$,
{\large M.B.~Gavela}~$^{b)}$,
{\large P.~Quilez}~$^{b)}$
\\
\vskip .7cm
{\footnotesize
$^{a)}$~
Institut f{\"u}r Theoretische Physik, Universit{\"a}t Heidelberg,\\
Philosophenweg 16, 69120, Heidelberg, Germany\\
\vskip .1cm
$^{b)}$~
Departamento de F\'isica Te\'orica and Instituto de F\'{\i}sica Te\'orica, IFT-UAM/CSIC,\\
Universidad Aut\'onoma de Madrid, Cantoblanco, 28049, Madrid, Spain
\vskip .5cm
}
\end{center}

\begin{abstract}
We determine the model-independent component of the couplings of axions to electroweak gauge bosons, induced by the minimal coupling to QCD inherent to solving the strong CP problem.
The case of the invisible QCD axion is developed first, and the impact on $W$ and $Z$ axion couplings is discussed.  The analysis is extended next to the generic framework of heavy true axions and low axion scales, corresponding to scenarios with enlarged confining sector. The mass dependence of the coupling of heavy axions to photons, $W$ and $Z$ bosons is determined. Furthermore, we perform a \emph{two-coupling-at-a-time} phenomenological study where the gluonic coupling together with individual gauge boson couplings are considered. In this way, the regions excluded by experimental data  for the axion-$WW$, axion-$ZZ$ and axion-$Z\gamma$ couplings are determined and analyzed together with the usual photonic ones.
The phenomenological results apply as well to ALPs which have anomalous couplings to both QCD and the electroweak bosons. 
\end{abstract}

\vskip 1cm

\begin{minipage}[l]{.9\textwidth}
{\footnotesize
\begin{center}
\textit{E-mail:} 
\email{alonso@thphys.uni-heidelberg.de},
\email{belen.gavela@uam.es},
\email{pablo.quilez@uam.es}
\end{center}}
\end{minipage}

\pagebreak
\tableofcontents

\pagebreak
%
%
\section{Introduction}
\label{Sect:Intro}
The term ``axion'' denotes  any (pseudo){Nambu-Goldstone boson (p{N}GB) of a global chiral $U(1)$ symmetry which is exact at the classical level but is explicitly broken only by  anomalous couplings to the field strength of a confining group.\footnote{The SM $\eta'$ is excluded from this definition since the associated $U(1)_A$ symmetry  is {also explicitely} broken by the non-zero quark masses.} Axions are the characteristic byproduct of solutions to the strong CP problem based on an anomalous $U(1)$ axial symmetry, usually called Peccei-Quinn (PQ) symmetry~\cite{Peccei:1977hh}.  Whatever its mass, an axion necessarily has anomalous gluonic couplings and it only deserves that name if its presence solves the strong CP problem. When the number of axions in a given theory outnumbers the total number of distinct instanton-induced scales other than QCD, one (or more) light axions remain.  
 
In scenarios with an axion $a$ coupled only to QCD, two states of the spectrum have anomalous couplings, the axion and the $\eta'$. Only one eigenstate can then acquire  a mass of the order $\Lambda_{QCD}$: this is identified as the physical $\eta'$, while the axion would remain massless in the limit of vanishing quark masses. The latter are relevant, though, and their impact on the mixing of pseudoscalars 
results in an axion mass given by~\cite{Weinberg:1977ma,Georgi:1986df}
\begin{equation}
m_a^2 f_a^2 \sim m_\pi^2 f_\pi^2\,\frac{m_u\,m_d}{(m_u+m_d)^2}\,,
\label{invisiblesaxion}
\end{equation}
where $m_\pi, f_\pi, m_u, m_d$ denote the pion mass and decay constant, and the up  and down quark masses, respectively. The axion scale to which all axion couplings are inversely proportional is denoted by $f_a$. The expression in Eq.~(\ref{invisiblesaxion}) is characteristic of all usual ``invisible'' axion constructions,  characterised by QCD being the only confining gauge group. Paradigmatic examples are the DFSZ and the KSVZ models~\cite{Kim79p103,Shifman80p493,Zhitnitsky:1980tq,Dine:1981rt} and their variations. These invisible axions are also called QCD axions.

 The main phenomenological constraints for QCD axions are obtained from their couplings to photons, gluons and fermions.  
  For instance, the effective coupling to photons $g_{a\gamma\gamma}$ is defined as   
\beq
 \delta\L \supset   \frac{1}{4}g_{a\gamma\gamma} \,a\,F_{\m\n}\tilde{F}^{\m\n}\,,
 \label{gagammagamma-def}
 \eeq
\beq\label{agammagamma_coupling}
g_{a\gamma\gamma} = -\frac{1}{2\pi f_a} \alphaem \left( \frac{E}{N} -1.92(4)\right)\,,
\eeq
where $F_{\m\n}$ is he electromagnetic field strength, $\alpha_{em}$ the corresponding fine structure constant, and $E$ and $N$ denote respectively the electromagnetic and color anomaly coefficients.  The ratio $E/N$ is model-dependent and has been recently computed in Refs.~\cite{DiLuzio:2017ogq,DiLuzio:2016sbl} for different representations of exotic fermions in standard ``invisible axion'' models. This ratio contains all direct loop contributions of PQ and electrically charged fermions, includ{ing} the up and down {and other SM} quarks if PQ charged. The second term in the parenthesis in Eq.~(\ref{agammagamma_coupling}) is instead model-independent.\cite{Kaplan85p215,Georgi:1986df,SREDNICKI1985689,Bardeen:1986yb,diCortona:2015ldu} It stems from the mixing of the axion with the $\eta'$ pseudoscalar -as both couple to  $G_{\mu\nu}\tilde G^{\mu\nu}$ with $G_{\mu\nu}$ being the QCD field strength- and also with the pion through the $\eta'$-pion mixing due to the up-down quarks mass difference. Obviously, such model-independent mixings are relevant for axions lighter than the QCD confining scale, while if the axion is much heavier this effect should become negligible. 

  Light enough axions {(that is, below $\mathcal{O}(100\,\mathrm{MeV})$)} can participate in astrophysical phenomena.~\cite{Raffelt:2006cw,Chang:2018rso,Irastorza:2018dyq} The constraints that follow from their non-observation in photonic processes lead to  very high values for {the decay constant},  $f_a>10^8$ GeV. It follows then from Eq.~(\ref{invisiblesaxion}) that  $m_a \le 10^{-5}$ eV for QCD axions.
 
Nevertheless, in specific QCD axion models the coupling to photons may be suppressed~\cite{DiLuzio:2016sbl, DiLuzio:2017pfr} and moreover large uncertainties may hover over the purely hadronic constraints.\cite{Alves:2017avw} It is thus important to analyze the axion couplings to {other} electroweak gauge bosons, as they may become the {phenomenologically} dominant couplings in certain regions of the parameter space for those models. Couplings of axions and also of axion-like particles (ALPs) to heavy gauge bosons are increasingly explored~\cite{Jaeckel:2015jla,Mimasu:2014nea,Bauer:2017ris,Bauer:2018uxu,Brivio:2017ije,Craig:2018kne} in view of present and future collider data, and also in view of rare meson decay data. For instance, in addition to LHC-related signals, recent works~\cite{Izaguirre:2016dfi,Freytsis:2009ct} consider the one-loop impact of $aWW$ couplings on rare meson decays.

   Here we first determine the model-independent components of the mixing of QCD axions with electroweak gauge bosons, which result from the mixing of the axion with the pseudoscalar mesons of the SM. In other words, we determine  the equivalent of the 1.92 factor in the photonic coupling in Eq.~(\ref{agammagamma_coupling}), for the couplings of the axion to $W$ and $Z$ gauge bosons. A chiral Lagrangian formulation will be used for this purpose, determining the leading-order effects.  The heavy electroweak gauge bosons will be introduced in that Lagrangian as external --classical-- sources. Our results  should impact the analyses for light axions of theories which solve the strong CP problem.  They are novel and relevant in particular whenever the axion is lighter than the QCD confining scale and is on-shell  in either low-energy or high-energy experiments.  They also impact the comparison between the data taken at experiments at low and high-energy. For instance, a null result in NA62 data for $K\rightarrow \pi a$ does not imply the absence of a signal at high energy in an accelerator such as that from off-shell axions at LEP or at a collider. This is because model-independent contributions are present at the low momenta dominant in rare decays (and cancellations may then take place), while at high energies they are absent. 
    
 As a second step, we will extend the analysis to heavy axions which solve the strong CP problem. Axions much heavier than $\Lambda_{QCD}$  and with low axion scales are possible within 
 dynamical solutions to the strong CP problem, at the expense of enlarging the confining sector of the Standard Model (SM) beyond QCD.\cite{Rubakov:1997vp,Fukuda:2015ana,Berezhiani:2000gh,Hsu:2004mf,Hook:2014cda,Chiang:2016eav,Dimopoulos:2016lvn,Gherghetta:2016fhp,Kobakhidze:2016rwh,Agrawal:2017ksf,Agrawal:2017evu,Gaillard:2018xgk} These theories introduce  a second and large instanton-induced scale $\Lambda'\gg \Lambda_{QCD}$ to which the axion also exhibits anomalous couplings, resulting in the bulk of its large mass.  Precisely because the axion mass typically lies well above the MeV regime, these heavy axions avoid the stringent astrophysical and laboratory constraints and present and future colliders may discover them. The transition  between the light and heavy axion regime will be explored for the coupling of the axion to the photon and to the electroweak gauge bosons.
 
 Finally, the phenomenological {part of the analysis} will be carried out on a ``two-coupling-at-a-time'' basis: it will take into account the simultaneous presence of a given electroweak coupling and the axion-gluon-gluon anomalous coupling (essential to solve the strong CP problem). 
  For the analysis of data,  for the first time  the experimentally excluded areas for the EW couplings  $g_{aWW}$, $g_{aZZ}$ and $g_{a\gamma Z}$ will be identified and depicted separately, besides  the customary ones for the $g_{a\gamma\gamma}$ coupling. Furthermore, the relations among the exclusion regions stemming from electroweak gauge invariance will be determined and exploited. 
  Model predictions will be illustrated over the experimental parameter space.
 
 Aside from the main focus of the paper on true axions, 
  our analysis applies to and calls for a timely extension of the ALP parameter space.  Very interesting bounds on ALPs from LEP and LHC~\cite{Mimasu:2014nea,Jaeckel:2015jla,Knapen:2016moh,Brivio:2017ije,Bauer:2017ris,Craig:2018kne,Bauer:2018uxu,CidVidal:2018blh}  assume often just one electroweak coupling for the axion, and no gluonic coupling. The path to consider any two (or more) couplings at a time will change the experimental perspective on ALPs. 
  
What is the difference between a heavy axion and an ALP with both anomalous electroweak and  gluonic couplings? The key distinction is that the former stems from a solution to the strong CP problem while a ``gluonic ALP'' may not.  Both exhibit anomalous couplings to QCD and in both cases there is an external source of mass besides that induced by QCD instantons and mixing. {However}, for a true heavy axion that extra source of mass  does not induce a shift of the $\theta$ parameter outside the CP conserving minimum (and thus the solution to the strong CP problem is preserved), while for a generic gluonic ALP {such a shift may be induced}. Nevertheless, this {important distinction} is not {directly} relevant for this work, as the novel aspects and phenomenological analysis developed below are valid for both  true heavy axions which solve the strong CP problem and for gluonic ALPs. To sum up, all results below for heavy axions apply directly to gluonic ALPs as well.  In addition, the conclusions based purely on EW gauge invariance have an even larger reach: they hold for all type of axions and for generic ALPs (that is, ALPs with or without gluonic couplings).

 The structure of the paper can be easily inferred from the Table of Contents.

 \section{ The Lagrangian for the QCD axion}
\label{Sect:2pointF_linear}

Without loss of generality, the axion couplings can be encoded in a model-independent way {in} an effective Lagrangian. At leading order in inverse powers of the scale $f_{\rm{PQ}}$ at which the global PQ symmetry is broken, and at energies above the electroweak (EW) scale, it reads
\beq
\LL_\text{eff}=\LL_\text{SM}\,+\frac{1}{2}(\partial_\mu \hat{a})(\partial^\mu \hat{a})\, + \delta\LL_a\,, 
\label{Lbosonic-lin}
\eeq
where $\hat{a}$ denotes the axion eigenstate at energies well above the confinement scale and $\L_\text{SM} $ is the SM Lagrangian,
 \begin{eqnarray} \label{LSM_lin}
\L_\text{SM} \supset 
&- &\fourth G^a_{\mu\nu}G^{a\mu\nu} - \fourth W^a_{\mu\nu}W^{a\mu\nu} - \fourth B_{\mu\nu}B^{\mu\nu}
+ D_\mu\Phi^\dag D^\mu\Phi \nonumber \\ 
&+&\sum_{\psi} i\bar{\psi}\slashed{D}\psi-\left(\bar{Q}_L\,\cY_D\,\Phi D_R+\bar{Q}_L\,\cY_U\,\tilde\Phi U_R+\bar{L}_L\,\cY_E\,\Phi E_R+\hc\right)\,.
\end{eqnarray}
In this expression, $W_{\mu\nu}$ and $B_{\mu\nu}$ are respectively the $SU(2)_L$ and $U(1)_Y$ gauge field strengths, while $\Phi$, $Q_L$, $L_L$, $U_R$, $D_R$ and $E_R$ respectively encode the  Higgs doublet, the fermion and lepton doublets, the vectors of right-handed up-type, down-type quarks and right-handed charged leptons (neutrino masses will be disregarded),  with $\psi$ running over all those type of fermions and  $\tilde{\Phi}\equiv i\s^2\Phi^*$.  
 The three $3\times 3$ matrices in flavour space $\cY_D$, $\cY_U$ and $\cY_E$ encode the Yukawa couplings for down quarks, up quarks and charged leptons, respectively.   
For the effective axion couplings, we first focus on the anomalous couplings of the axion and their impact on the pseudoscalar mass matrix. 

We will work in the basis in which the only PQ-breaking operators in the Lagrangian, $\delta\LL_a^{\cancel{PQ}}$, are  the anomalous couplings of axions to gauge bosons. This choice is allowed by the reparametrization invariance of the effective axion Lagrangian (see Appendix~\ref{App: general matrix}). 
 The CP-conserving and PQ-violating next-to-leading order (NLO) corrections due to axion physics then read~\footnote{The derivative operators also present in the most general basis~\cite{Georgi:1986df,Choi:1986zw, Salvio:2013iaa,Brivio:2017ije} are PQ invariant and thus not shown. } 
  \begin{equation}
 \delta\LL_a^{\cancel{PQ}}\,=N_0\,\O_{\tilde{G}}+\,L_0\,\O_{\tilde{W}}+P_0\,\O_{\tilde{B}}\,,
 \label{general-NLOLag-lin}
\end{equation}
 with $\O_{\tilde{G}}$, $\O_{\tilde{W}}$ and  $\O_{\tilde{B}}$ denoting the anomalous axion couplings to gluons, $SU(2)_L$ and $U(1)_Y$ gauge bosons, respectively, 
 \begin{eqnarray}
 \O_{\tilde{G}} &\equiv&-\frac{\alpha_s}{8\pi}\,G^a_{\mu\nu}\tilde{G}^{a\mu\nu}\dfrac{\hat{a}}{f_{\rm{PQ}}}\,,\label{AGtilde}\\
 \O_{\tilde{W}} &\equiv&-\frac{\alpha_W}{8\pi}\WWd^a\tilde{W}^{a\mu\nu}\dfrac{\hat{a}}{f_{\rm{PQ}}}\,,\label{Wtilde}\\
\O_{\tilde{B}} &\equiv&-\frac{\alpha_B}{8\pi}\,\BBd\tilde{B}^{\mu\nu}\dfrac{\hat{a}}{f_{\rm{PQ}}}\,,\label{ABtilde}
\label{OaPhi}
\end{eqnarray}
where  $\alpha_s$, $\alpha_W$ and $\alpha_B$ denote respectively the fine structure constants for the QCD, $SU(2)_L$ and $U(1)$ gauge interactions,
and $N_0$, $P_0$ and $L_0$ are dimensionless operator coefficients.  Customarily, the Lagrangian in Eq.~(\ref{general-NLOLag-lin}) is rewritten as 
\begin{align}
\delta\LL_a\,= 
  & \fourth g_{agg}^0 \,\hat{a}\, G\tilde{G} +  \fourth g_{aWW}^0 \,\hat{a}\, W\tilde{W}  + \fourth g_{aBB}^0 \,\hat{a} \,B\tilde{B}\,,
 \label{Eq:maldicion}
\end{align}
where the Lorentz indices of the field strengths are implicit from now on and 
\begin{align}
g_{agg}^0 \equiv-\frac{1}{2\pi f_{\rm PQ}} \,{\alpha_s}\,N_0\,,\qquad
g_{aWW}^0 \equiv-\frac{1}{2\pi f_{\rm PQ}} \alpha_W \, {L_0}\,,\qquad
g_{aBB}^0 \equiv-\frac{1}{2\pi f_{\rm PQ}} {\alpha_B} \,{P_0} \,.
\label{gBB}
\end{align}
The model-dependent group theoretical factors can  be generically written in terms of the fermionic PQ charges $\mathcal{X}_{\,i}$   as
\begin{eqnarray}
N_0&=&\sum_{i=\rm{heavy}} 2\,\mathcal{X}_{\,i} \,T(R^{SU(3)}_{i})\,, \nn\\
L_0&=&\,\sum_{i=\rm{heavy}}2\, \mathcal{X}_i  T(R^{SU(2)}_{i})\,, \nn\\
P_0&=&\,\sum_{i=\rm{heavy}} 2\,\mathcal{X}_i  \, Y_i^2\,, \label{NE}
\end{eqnarray}
where  $\mathcal{X}_i $ is the difference between the right-handed and left-handed PQ charge{}s:\footnote{Obviously, only left-handed quarks may contribute to $L_0$; in any case, it is always possible to work in a convention in which only left-handed quarks are PQ charged. }
\begin{equation}
\mathcal{X}_i = \mathcal{X}_{Li}-\mathcal{X}_{Ri}\,.
\end{equation}
In Eq.~(\ref{NE}), $T(R^{SU(3)}_{i})$ and $T(R^{SU(2)}_{i})$  are respectively the Dynkin indices of the fermionic representation $R_i$  under QCD and $SU(2)_L$, and $Y_i$ denotes the hypercharge.  
  The sum over ``heavy'' fermions and the subscript $0$ indicate that the contribution to the anomalous couplings from all exotic heavy quarks, and/or heavy SM quarks ($s,\,c,\,b,\text{ and }t$ quarks) if PQ charged, is encoded in the $N_0$, $L_0$ and $P_0$ operator coefficients. That is,   the possible contribution of the SM first generation up ($u$) and down ($d$) quarks is not included in those coefficients. Indeed, for models in which they have PQ charges  an additional PQ-invariant term must be considered, replacing  the $u$ and $d$ Yukawa couplings in Eq.~(\ref{LSM_lin}) by:
\begin{equation}
 \delta\LL_a^{{PQ}}\,=-\bar{Q}_{1L}\,\cY_d\,\Phi\, d_R\,e^{i\,\mathcal{X}_d\,\hat a/f_{\text{PQ}}} -\bar{Q}_{1L}\,\cY_u\,\tilde\Phi\, u_R\,e^{i\,\mathcal{X}_u\,\hat a/f_{\text{PQ}}} 
 +\hc \,, 
\label{general-NLOLag-c2}
\end{equation}
which assumes as convention that the axion transforms under the PQ symmetry as $a\to a + \fpq$. 
The $\mathcal{X}_{u,d}$ dependence in Eq.~(\ref{general-NLOLag-c2}) will be shown below to generate extra contributions to the physical anomalous couplings. 
 In this equation $Q_1$ denotes the first family doublet, and flavour-mixing effects as well as leptonic couplings are omitted from now on for simplicity. In all equations above, color and $SU(2)_L$ indices are implicit and the QCD $\theta$ angle has been removed from the Lagrangian via the PQ symmetry.    
 
 Among the most general set of purely derivative operators, additional couplings could also be considered, e.g. 
 \begin{equation}
 \delta\LL_{a, \text{deriv}}^{{PQ}}\,=-\frac{\de_\mu \hat{a}}{f_{\text{PQ}}}  \left(\bar{Q}_{L} \,\gamma_\mu\, c_{1}^{Q}\, Q_{L}\,
  +\,\bar{U}_{R} \,\gamma_\mu\,  c_{1}^{U}\,U_{R}\,
  +\,\bar{D}_{R} \,\gamma_\mu\, c_{1}^D\,D_{R}\,
  \right)\,,
\label{general-NLOLag-c1}
\end{equation}
where $c_{1}^{Q}$, $c_{1}^{U}$ and $c_{1}^{D}$ are matrices of arbitrary coefficients in flavour space.  Nevertheless,  the reparametrization invariance of the Lagrangian~\cite{Kim10p557} allows to work in a basis in which these terms (which would seed pseudoscalar kinetic mixing) are absent and their impact is transferred to other axionic couplings.\footnote{
In App.~\ref{App: general matrix} it will be explicitly shown that they do not have physical impact on mixing. Note that  possible flavour non-diagonal couplings are not considered.}
From now on they will be disregarded all through the analysis on pseudoscalar mixing. 
In summary, the Lagrangian to be analyzed  below when considering mixing effects  reads
\begin{equation}
\delta\LL_{a} \,=\, \delta\LL_a^{\cancel{PQ}}\,\,+\,\delta\LL_{a}^{{PQ}} \,.
\label{Eq: Lag pq nopq}
\end{equation}
 
\subsubsection*{Below EW symmetry breaking and above confinement}
After electroweak symmetry breaking but before QCD confinement, the  
effective Lagrangian in Eq.~(\ref{Eq: Lag pq nopq})  
results in
\begin{align}
\delta\LL_a\,= 
 -&\,\bar{u}_L\, m_u\, u_R \,e^{i\,\mathcal{X}_u\,\hat a/f_{\text{PQ}}} \, -
 \, \bar{d}_L\, m_d\, d_R \,e^{i\,\mathcal{X}_d\,\hat a/f_{\text{PQ}}}\, +\, \hc \nn\\
 + & \fourth g_{agg}^0 \,\hat{a}\, G\tilde{G} +  \fourth g_{a\gamma\gamma}^0 \,\hat{a}\, F\tilde{F}  + \fourth g_{aWW}^0 \,\hat{a} \,W\tilde{W} + \fourth g_{aZZ}^0 \,\hat{a}\, Z\tilde{Z}  + \fourth g_{a\gamma Z}^0 \,\hat{a}\, F\tilde{Z}\,,
 \label{Eq:Lgeneral-after}
\end{align} 
where 
 \begin{align}
g_{agg}^0 & =-\frac{1}{2\pi f_{\rm PQ}} \,{\alpha_s}\,N_0\,\label{gagluon}\\
g_{a\gamma\gamma}^0 &= -\frac{1}{2\pi f_{\rm PQ}} \alphaem \, {E_0}\,,\\
g_{aWW}^0 &= -\frac{1}{2\pi f_{\rm PQ}} \frac{\alphaem}{s_w^2} \,{L_0} \,,\\
g_{aZZ}^0 &= -\frac{1}{2\pi f_{\rm PQ}} \frac{\alphaem}{s_w^2c_w^2} \, Z_0 \,,\\
g_{a\gamma Z}^0 &= -\frac{1}{2\pi f_{\rm PQ}} \frac{\alphaem}{s_wc_w} \, {2R_0}\,,
\label{gXX0}
\end{align}
In these equations $s_w$ and $c_w$ denote the sine and cosine of the Weinberg mixing angle and $\alpha_{em}= \alpha_{W} c_w^2=   \alpha_{B} s_w^2$. 
 
  For models in which the  the first generation of SM quarks are not PQ charged, $\mathcal{X}_{u,d}=0$. When those quarks are instead charged under PQ, their contribution to the anomalous gauge couplings has to be included in the group theory factors,  which are replaced by 
 \begin{equation}
N =N_0 + N_{u,d}\,, \qquad  \qquad E =E_0 + E_{u,d}\,,
\qquad  \qquad L =L_0 + L_{u,d}\,,
\label{N0}
\end{equation}
  \begin{equation}
Z =Z_0 + Z_{u,d}\,, \qquad  \qquad R =R_0 + R_{u,d}\,,
\label{Ntotal}
\end{equation}
as computed further below.  
 In all cases, only two among the four parameters $E$, $L$, $Z$ and $R$ are linearly independent,  because of gauge invariance, see  Eq.~(\ref{general-NLOLag-lin}), 
\begin{equation}
E \equiv L+P\,, \qquad Z\equiv Lc_w^4 + Ps_w^4\,, \qquad
R \equiv L c_w^2-P s_w^2\,.
\label{Eq:coupling-relations-gauge}
\end{equation}
A non-vanishing $N$ is the trademark of axion models which solve the strong CP problem, while the presence of the other couplings is  model-dependent.  It is customary to define the physical axion scale $f_a$ from the strength of the gluonic coupling: 
  \begin{equation}
   f_{a} \equiv \frac{f_{PQ}}{    N} \,.
      \end{equation}
\subsection{ The Lagrangian below the QCD confinement scale}
Three pseudoscalars mix once quarks are confined: the axion, the SM singlet  $\eta_0$ and the neutral pion $\pi_3$. 
 The $\pi_3$-$\eta_0$ mixing is due to the quark mass differences which break the global flavour symmetry. At leading order in the chiral expansion  and in the two quark approximation, the mass Lagrangian for the pions and $\eta_0$ reads 
 \begin{equation}
\mathcal{L}\supset B_0 \frac{f_{\pi}^2}{2}\, \rm{Tr}\left( \mathbf{\Sigma}\, M_q^\dagger + M_q\,\mathbf{\Sigma}^\dagger \right)\,,
\label{Lchiral}
\end{equation}
where $B_0$ can be expressed  in terms of the QCD quark condensate $\langle \bar{q} q\rangle$ as  $B_0 f_{\pi}^2=-2  \langle \bar{q} q\rangle$, 
and $M_q$ denotes the quark mass matrix,
\begin{align}\label{eq:ud_mass_matrix}
M_q=\left(
\begin{matrix}
m_u & 0\\
0 & m_d 
\end{matrix} 
\right)\,.
\end{align}
 The matrix of pseudoscalar fields can be written as 
\beq
\mathbf{\Sigma}(x)=\, \rm{exp}{[i(2\eta_0/(f_\pi\,\sqrt{2})\, \mathbf{1}]}\,\rm{exp}{[i\,\mathbf{\Pi}/f_\pi]}\, , \qquad 
\eeq
where the $\eta_0$ decay constant has been approximated by $f_\pi$ and 
\begin{align}
\mathbf{\Pi}\equiv\left(\begin{matrix}  \pi_3 & 
	\sqrt{2}\pi^+  \\ 
 \sqrt{2}\pi^-  &	  -\pi_3
	\end{matrix}\right)\,.
\end{align}
In the presence of the axion, the anomalous QCD current $G\tilde{G}$ couples to both the axion and the  $\eta_0$  fields and mixes them.  The two mixing sources combined result ultimately in an axion-pion mixing.  
 {For simplicity}, we will first consider the case with $\mathcal{X}_{u,d}=0$ in Eq.~(\ref{general-NLOLag-c2}), and afterwards the case $\mathcal{X}_{u,d}\ne 0$. 

\subsubsection{ SM light quarks not charged under PQ ($\mathbf{\mathcal{X}_{u,d}=0}$)}

A popular example of this class of models are KSVZ ones, in which only heavy exotic quarks are charged under PQ. 
For any model in which the $u$ and $d$ SM quarks are singlets of the PQ symmetry, their quark mass matrix in the basis here considered is that in Eq.~\eqref{eq:ud_mass_matrix}.
In this case $N\,=\, N^0$, as all contributions to the anomalous gluonic coupling are already included in $N^0$. The potential for the three pseudoscalars is, at first order in the pseudoscalar masses, 
\begin{align}
V=-B_0 f_{\pi}^2 \left[m_u\,\,\text{cos}\left(\frac{\pi_3}{f_\pi}+\frac{\eta_0}{f_\pi}\right)+m_d\,\text{cos}\left(\frac{\pi_3}{f_\pi}-\frac{\eta_0}{f_\pi}\right)\right]+ \frac{1}{2}\,K\,\left(2\frac{\eta_0}{f_\pi}+\frac{a}{f_a}\right)^2\,,
\label{potc20}
\end{align} 
 where the last term stems from the instanton potential with  $K\sim \Lambda_{QCD}^4$.\cite{DiVecchia:1980yfw,DiVecchia:2017xpu,DiVecchia2014}
The resulting mass matrix for the three neutral pseudoscalars is given by
\beq
M^2_{\{\pi_3,\,\eta_{0},a\}}=
\left(
\begin{array}{ccc}
B_0\,(m_u+m_d)  & \quad B_0\,(m_u-m_d)&0\\ 
B_0\,(m_u-m_d)  &  \quad 4K/
{f_\pi}+B_0(m_u+m_d) &\quad {2K}/({f_\pi f_a})\\ 
0 & {2 K}/({f_\pi f_a})  &  {K}/{f_a^2}
\end{array}  
\right)\,.\label{Eq:massmatrix}
\eeq
The diagonalization leads to the well-known expressions for the pseudoscalar mass terms:
\begin{align}
m^2_{\pi}\simeq B_0\,(m_u+m_d),\quad m^2_{\eta'}\simeq\frac{4K}{f^2_\pi}+B_0\,(m_u+m_d), \quad m^2_{a}\simeq \frac{f_\pi^2m_\pi^2}{f_a^2}\frac{m_um_d}{(m_u+m_d)^2}\,.
\label{masses}
\end{align}
 It follows from these expressions that  $K$ can be expressed in terms of the low-energy physical parameters as
\begin{equation}
K\simeq \frac{1}{4}\,(m_\eta^2 - m_\pi^2)\, f_{\pi}^2\,.
\end{equation}
The corresponding mixing matrix is given by
\begin{align}
\left(\begin{matrix}  1 & 
	\frac{f \left(m_{d} - m_{u}\right)}{2 f_{a} \left(m_{d} + m_{u}\right)}     & 
	\frac{f}{2 f_{a}} \\ 
-\frac{f \left(m_{d} - m_{u}\right)}{2 f_{a} \left(m_{d} + m_{u}\right)}     &	
	 1 & 
	 - \frac{ m_\pi^{2}}{m^2_{\eta'}}\frac{ \left(m_{d} - m_{u}\right)}{ \left(m_{d} + m_{u}\right)} \\ 
- \frac{f}{2 f_{a}}  & \frac{ m_\pi^{2}}{m^2_{\eta'}}\frac{ \left(m_{d} - m_{u}\right)}{ \left(m_{d} + m_{u}\right)} & 1\end{matrix}\right)\,,
\label{mixingmatrixnoc2}
\end{align}
or, equivalently, the mass eigenstates are given by
\begin{align}
a & \simeq \hat{a} +\theta_{a\pi}\,\pi_3 + \theta_{a\eta'}\,\eta_0\,,\label{aphys}\\
\pi^0 & \simeq \pi_3 + \theta_{\pi\,a}\,{a} + \theta_{\pi\eta'}\,\eta_0\,,\\
\eta' & \simeq   \eta_0+\theta_{\eta'\,a}\,{a} + \theta_{\eta'\,\pi}\,\pi_3 \,.
\label{mixings}
\end{align}
Here, all the mixing angles are assumed small and 
\begin{align}
&\theta_{a\pi}\simeq -\frac{f_\pi}{2f_a}\frac{m_d-m_u}{m_u+m_d}\,, \qquad &\theta_{a\eta'} \simeq -\frac{f_\pi}{2f_a}\,,\qquad &\qquad \theta_{\pi\eta'} \simeq \frac{ m_\pi^{2}}{m^2_{\eta'}}\frac{ \left(m_{d} - m_{u}\right)}{ \left(m_{d} + m_{u}\right)}\,,\label{thetas-1}\\
&\theta_{\pi\,a}\simeq -\theta_{a\pi}\,,    \qquad & \theta_{\eta'\,a}\simeq -\theta_{a\eta'}\,, \qquad &\qquad \theta_{\eta'\,\pi}\simeq -\theta_{\pi\eta'}\,.
\label{angles}
\end{align}
Only the leading terms  for each mixing entry have been kept in Eqs.~(\ref{mixingmatrixnoc2})-(\ref{angles}). The complete matrix to first order in $1/f_a$ and in quark masses --that is, $\mathcal{O}(m_\pi ^2/m_\eta^2)$--  can be found in Appendix~\ref{masscomplete}: the impact of the extra terms may be comparable to that of next-to-leading operators in the chiral expansion and will thus not be retained here. 

The results in Eqs.~(\ref{aphys})-(\ref{angles}) illustrate that the physical low-energy axion eigenstate  acquires $\pi_3$ and $\eta_0$ components and thus 
inherits their couplings   to {\it all gauge bosons}, weighted down by their mixing with the axion. These results apply to any physical process in which the axion is on-shell and the axion mass is lighter than the confinement scale.  

We are interested in identifying the model-independent contributions in the coupling to the electroweak gauge bosons for light axions and for the SM light pseudoscalars.  We will first recover in our basis the customary axion-photon couplings, to set the framework.

\subsubsection*{Axion-photon coupling} 
The  physical $g_{a\gamma\gamma}$ is given by 
\beq
g_{a\gamma\gamma} = g_{a\gamma\gamma}^0 + \theta_{a\pi}\, g_{\pi \gamma\gamma}+ \theta_{a\eta'}\,g_{\eta' \gamma\gamma}\,,
\label{gagamma}
\eeq
where the last two terms are the contributions induced by the model-independent  axion-pion and axion-$\eta'$  QCD mixing. 
Denoting by $q_u$ and $q_d$ the electric charges of the up and down quarks, respectively,  the photonic couplings of the SM light pseudoscalars  are given by
\begin{align}
&g_{\pi \gamma\gamma} \equiv -\frac{3\,\alpha}{\pi\,f_\pi} \big(q_u^2-q_d^2\big)\,, 
&g_{\eta' \gamma\gamma} \equiv -\frac{3\,\alpha}{\pi\,f_\pi} \big(q_u^2+q_d^2\big)\,.
\end{align}
For the present case with ${\mathcal{X}_{u,d}=0}$, that is $E_0=E$ and  $N_0=N$, it follows  that 
\begin{equation}
g_{a\gamma\gamma} = g_{a\gamma\gamma}^0 + \frac{\alpha}{2\pi f_a}\left(6\,  \frac{q_d^2\,m_u+q_u^2\,m_d}{m_u+m_d}\right)\,,
\end{equation}
resulting in the well-known expression
\begin{equation}
g_{a\gamma\gamma} =  - \frac{\alpha}{2\pi f_a}\left(\frac{E}{N}-\frac{2}{3}\,\frac{m_u+4\,m_d}{m_u+m_d}\right)\,,
\end{equation}
which is valid to first order in chiral perturbation theory.

\subsubsection{ SM light quarks charged under PQ ($\mathbf{\mathcal{X}_{u,d}\ne0}$)}
The quark mass matrix in Eq.~(\ref{Lchiral}) is to be replaced by
\begin{align}
M_q=\left(
\begin{matrix}
m_u & 0\\
0 & m_d 
\end{matrix} 
\right)
\left(
\begin{matrix}
e^{i\,\mathcal{X}_u\,\hat a/f_{\text{PQ}}} & 0\\
0 & e^{i\,\mathcal{X}_d\,\hat a/f_{\text{PQ}}} 
\end{matrix} 
\right)\,. 
\end{align}
 The potential in Eq.~(\ref{potc20}) is now generalized to
\begin{multline}
V=-B_0 f_{\pi}^2 \left[
m_u\,\text{cos}\left(\frac{\pi_3}{f_\pi}+\frac{\eta_0}{f_\pi}-\,\mathcal{X}_u\,\frac{\hat a}{\fpq}\right)+
m_d\,\text{cos}\left(\frac{\pi_3}{f_\pi}-\frac{\eta_0}{f_\pi}-\,\mathcal{X}_d\,\frac{\hat a}{\fpq}\right)\right]
\,\\
+\,\frac{1}{2}\,K\,\left[2\frac{\eta_0}{f_\pi}+\,N_0\frac{\hat a}{\fpq}\right]^2\, 
\label{potc2}\,,
\end{multline} 
 resulting in a pseudoscalar squared mass matrix $M^2_{\{\pi_3,\,\eta_{0},a\}}$ which takes the form
\begin{equation}
\scriptstyle M^2_{\{\pi_{3},\,\eta_{0},a\}}=
\left(
\begin{array}{ccc}
\scriptstyle B_0\,(m_u+m_d)  & \scriptstyle \quad B_0\,(m_u-m_d)&
\scriptstyle -B_0 \frac{f_\pi}{\fpq} \big(m_u\,\mathcal{X}_u-m_d\,\mathcal{X}_d\big)\\ 
\scriptstyle B_0\,(m_u-m_d)  &  \scriptstyle \quad \frac{4K}{f_\pi}+B_0(m_u+m_d) &
\scriptstyle \frac{2\,N_0 K}{f_\pi \fpq}+B_0 \frac{f_\pi}{\fpq} \big(m_u\,\mathcal{X}_u+m_d\,\mathcal{X}_d\big)\\ 
\scriptstyle -B_0 \frac{f_\pi}{\fpq} \big(m_u\,\mathcal{X}_u-m_d\,\mathcal{X}_d\big) & 
\scriptstyle \frac{2\,N_0 K}{f_\pi \fpq}+B_0 \frac{f_\pi}{\fpq} \big(m_u\,\mathcal{X}_u+m_d\,\mathcal{X}_d\big)  &  
\scriptstyle \frac{N_0^2\,K}{\fpq^2} +B_0 \frac{f_\pi^2}{\fpq^2} \big(m_u\,\mathcal{X}_u^{\,2}+m_d\,\mathcal{X}_d^2\big)
\end{array}  
\right).\label{Eq:GeneralMassmatrixPQ}
\end{equation}
The mixing angles in Eqs.~(\ref{thetas-1})-(\ref{angles}) still hold but for the pion-axion mixing which is now given by~\footnote{This expression for the axion-pion mixing agrees with the result of Ref.~\cite{Alves:2017avw} for the case where the only PQ charged fermions are the up and down quarks, i.e. $N_0=0$. }
\begin{align}
&\theta_{a\pi}\simeq -\frac{f_\pi}{2\fpq}\left(\mathcal{X}_d-\mathcal{X}_u+\big(N_0+\mathcal{X}_d+\mathcal{X}_u\big)\frac{m_d-m_u}{m_u+m_d}\right)\,.
\end{align} 
The coefficient in front of the mass-dependent term in this equation coincides with the strength of the physical gluonic couplings,\footnote{As expected from the triangle diagram, all fermions (including the up and down quarks) run in the loop and contribute to $N=N_0+\sum_{u,\,d} 2\, \mathcal{X}\,T(R)=N_0+\mathcal{X}_d+\mathcal{X}_u$.} given by
\begin{equation}
N=\big(N_0+\mathcal{X}_d+\mathcal{X}_u\big)\,,
\label{N}
\end{equation}
and in consequence 
\begin{align}
&\theta_{a\pi}\simeq -\frac{f_\pi}{2Nf_a}\left(\mathcal{X}_d-\mathcal{X}_u+N\frac{m_d-m_u}{m_u+m_d}\right)\,.
\end{align}

The expressions for the mass of the physical pion, $\eta'$ and axion are the same than those in  Eq.~(\ref{masses}).
\subsubsection*{Axion-photon coupling}
For the case in which the up and down quarks are charged under PQ, ${\mathcal{X}_{u,d}\ne 0}$, $N$ is given by Eq.~(\ref{N}) resulting in 
\begin{align}
g_{a\gamma\gamma} & = \,g_{a\gamma\gamma}^0 - \frac{\alpha}{2\pi f_a}\left(\frac{E_{u,\,d}}{N} -\frac{2}{3}\frac{m_u+4\,m_d}{m_u+m_d}\right)\,,\end{align}
where
\begin{equation}
E_{u,\,d}\, = \, 6\,\mathcal{X}_u\, q_u^2\, + \,6\,\mathcal{X}_d\, q_d^2\,. 
 \end{equation}
 In consequence
 \begin{equation}
g_{a\gamma\gamma} = -\frac{\alpha}{2\pi f_a}\left(\frac{E_0}{N}+\frac{E_{u,\,d}}{N} -\frac{2}{3}\frac{m_u+4\,m_d}{m_u+m_d}\right) = -\frac{\alpha}{2\pi f_a}\left(\frac{E}{N} -\frac{2}{3}\frac{m_u+4\,m_d}{m_u+m_d}\right)\,.
\end{equation}
In summary,  the most general mass matrix leads to the same expression than for the case $\mathcal{X}_{u,d}=0$  in Eq.~(\ref{gagamma}) if taking into account  in $E$ also the contribution of the up and down quarks.

\subsection{Axion couplings to EW gauge bosons}
\label{axion-EW}
The description in terms of the effective chiral Lagrangian is only appropriate for energies/momenta not higher than the cutoff of the effective theory, $4\pi f_\pi$, that is, the QCD scale as set by the nucleon mass. In this context, the $W$ and $Z$ bosons can be considered as external currents that couple to a QCD axion whose energy/momentum is not higher than $\Lambda_{QCD}$, for instance a light enough on-shell QCD axion. In other words, the $W$ and $Z$ bosons enter the effective chiral Lagrangian as classical sources, alike to the treatment of baryons in the effective chiral Lagrangian. 

We extend here the results of the previous section to the interactions of axions with SM heavy gauge bosons.  Eq.~(\ref{gagamma}) is thus generalized for any pair of electroweak gauge bosons $X$, $Y$,  
\beq
g_{aXY} = g_{aXY}^0 + \theta_{a\pi}\, g_{\pi XY}+ \theta_{a\eta'}\,g_{\eta' XY}\,. 
\eeq
Indeed, all axion couplings to EW bosons receive a model-independent component due to the pion-$\eta'$-axion mixings, in the regime in which the involved energy/momenta are smaller or comparable to the confinement scale. 
 A relevant point when computing the  couplings of the QCD axion to heavy EW bosons is the fact that, after confinement,   a new type of $SU(2)_L$-breaking effective interaction of the form 
\begin{equation}
 {\hat{a}}\,\WWd^3\tilde{B}^{\mu\nu}\,,
 \label{aBW}
\end{equation}
 is present  in addition to those in Eqs.~(\ref{AGtilde})-(\ref{ABtilde}). It
 stems via axion-pion coupling from the $ {{\pi_a}} \WWd^a\tilde{B}^{\mu\nu}$ effective interaction.
 The details of the computation can be found in App.~\ref{appendix-anomalous}.  
 In terms of the physical photon, $Z$ and $W$, the  interaction Lagrangian for the QCD axion is then given by
\begin{align}
 \delta\LL_a^{gauge}\,= & \fourth g_{agg} \,a \,G\tilde{G} + \fourth g_{aWW} \,a \,W\tilde{W} + \fourth g_{aZZ} \,a\, Z\tilde{Z} + \fourth g_{a\gamma\gamma} \,a\, F\tilde{F} + \fourth g_{a\gamma Z} \,a\, F\tilde{Z}\,, 
 \label{LunderConf}
\end{align} 
where 
\begin{align}
g_{agg} &= -\frac{1}{2\pi f_a} \alpha_s \,, \label{gagg}\\
g_{a\gamma\gamma} &= -\frac{1}{2\pi f_a} \alphaem \left( \frac{E}{N} - \frac{2}{3}\frac{m_u+4m_d}{m_u+m_d} \right)\,, \label{gagamma}\\
g_{aWW} &= -\frac{1}{2\pi f_a} \frac{\alphaem}{s_w^2} \left( \frac{L}{N} - \frac{3}{4} \right)\,, \label{Eq:gaW}\\
g_{aZZ} &= -\frac{1}{2\pi f_a} \frac{\alphaem}{s_w^2c_w^2} \left( \frac{Z}{N} - \frac{11s_w^4+9c_w^4}{12} - \frac{s_w^2(s_w^2-c_w^2)}{2}\frac{m_d-m_u}{m_u+m_d} \right)\,, \label{Eq:gaZ}\\
g_{a\gamma Z} &= -\frac{1}{2\pi f_a} \frac{\alphaem}{s_wc_w} \left( \frac{2R}{N} - \frac{9c_w^2-11s_w^2}{6} - \frac{1}{2}(c_w^2-3s_w^2)\frac{m_d-m_u}{m_u+m_d} \right)\,.\label{Eq:gagammaZ}
\end{align}
Eq.~(\ref{gagamma}) is the known leading-order result~\cite{Kaplan85p215,diCortona:2015ldu} for the photonic couplings of the QCD axion, which holds in all generality for on-shell axions lighter than the QCD confinement scale. The contributions in Eqs.~(\ref{Eq:gaW})-(\ref{Eq:gagammaZ}) are new and extend that result to the couplings  of heavy gauge bosons in the appropriate energy range.  Indeed, the last term in the parenthesis for each of these couplings encodes the impact of the mixing of the axion with the pion and $\eta'$, see Eqs.~(\ref{gagluon})-(\ref{gXX0}) for comparison with the unmixed case. These are model-independent contributions, valid for any QCD axion, i.e.,  for any model in which the SM strong gauge group is the only confining force and thus the only source of an instanton potential for the axion. In other words, they are valid  for any axion whose mass and scale are related by Eq.~(\ref{masses}). Note that those corrections hold precisely because the axion mass is smaller than the confining scale of QCD, which is the regime in which the SM pseudoscalars lighter than the QCD scale are the physical eigenstates of the spectrum and the mixing effects are meaningful. 

Numerically, at leading order in the chiral expansion it follows that
\begin{align}
g_{a\gamma\gamma} &= -\frac{1}{2\pi f_a} \alphaem \left( \frac{E}{N} - 2.03\right)\,, \label{gagamma-number}\\
g_{aWW} &= -\frac{1}{2\pi f_a} \frac{\alphaem}{s_w^2} \left( \frac{L}{N} - 0.75 \right)\,, \label{gaW-number}\\
g_{aZZ} &= -\frac{1}{2\pi f_a} \frac{\alphaem}{s_w^2c_w^2} \left( \frac{Z}{N} - 0.52 \right)\,, \label{gaZ-number}\\
g_{a\gamma Z} &= -\frac{1}{2\pi f_a} \frac{\alphaem}{s_wc_w} \left( \frac{2R}{N} - 0.74  \right)\,.\label{gagammaZ-number}
\end{align}
The numerical value of the model-independent term in Eq.~(\ref{gagamma-number}) differs from the usual one~\cite{diCortona:2015ldu} of 1.92 in Eq.~(\ref{agammagamma_coupling}), as the latter includes higher order chiral corrections, a refinement out of the scope of this paper and left for future work. 

The model-independent results obtained here for the coupling of light QCD axions to the SM electroweak bosons may impact axion signals in rare decays in which they participate. For instance, in low-energy processes the axion could be be photophobic at low energies~\cite{DiLuzio:2017pfr} (or more generally, EW-phobic), in models in which the terms in parenthesis cancel approximately, unlike at higher energies at which the model-independent component disappears and only the model-dependence (encoded in $E/N$, $M/N$, $Z/N$ and $R/N$) is at play.

\subsubsection*{Gauge invariance}
As it was already enforced in Eq.~(\ref{Eq:coupling-relations-gauge}), the couplings of the axion to the EW gauge bosons are not independent as a consequence of gauge invariance.   Indeed,  all four couplings stem from the two independent effective operators in Eqs.~(\ref{Wtilde})  and (\ref{ABtilde}), plus that in Eq.~(\ref{aBW}) for a light QCD axion  ($m_a\ll\Lambda_{QCD}$).  In consequence,  three physical couplings can be independent among the set  $\{g_{a\gamma\gamma}\,, g_{aWW}\,,\,g_{aZZ}\,, g_{a\gamma Z}\}$, and the following relation must  hold:
\begin{align}
g_{aZZ}&= -\left(c_w^2+\frac{s_w^4}{c_w^2}\right) g_{aWW}+\frac{c_w^3}{s_w}g_{a\gamma\gamma} + \left(1+c_w^2+\frac{s_w^4}{c_w^2}\right) g_{a\gamma Z}\,.
\label{constrain-light}
\end{align}
 Note that this result does not depend on the details of any particular axion model; it is independent of the presence or absence of gluonic couplings and it thus applies in all generality for a light pseudoscalar with only anomalous electroweak  
 couplings. That is, it is also valid for generic ALPs which only have EW interactions. Furthermore, it suggests that it may be inconsistent to assume only one EW coupling: the minimum number of physical EW couplings for an axion or ALP is two.  
  The relation in Eq.~(\ref{constrain-light}) sets an avenue to oversconstrain the parameter space which is promising. It allows to use the better constrained EW couplings  to bound the fourth one.  
  
 \section{Beyond the QCD axion}
\label{sec:ExtraMass}
We discuss in this section the case of a ``heavy axion'': an axion whose mass is not given by the QCD axion expression in Eq.~(\ref{invisiblesaxion}) but receives instead extra contributions. We have in mind a true axion which solves the strong CP problem, for which the source of this extra mass does not  
 spoil the alignment of the CP conserving minimum. This is the case for instance of models in which the confining sector of the SM is enlarged involving a new force with a confining scale much larger than the QCD one~\cite{Rubakov:1997vp,Fukuda:2015ana,Berezhiani:2000gh,Hsu:2004mf,Hook:2014cda,Chiang:2016eav,Dimopoulos:2016lvn,Gherghetta:2016fhp,Kobakhidze:2016rwh,Agrawal:2017ksf,Agrawal:2017evu,Gaillard:2018xgk}. This avenue is of particular interest as it allows to consider heavy axions  and low axion scales (e.g. $\mathcal{O}$(TeV)), and still solve the SM strong CP problem. The axion mass can then expand a very  large range of values.    
It can become much larger than the EW scale 
 or, conversely, be in the GeV range or lower. For the purpose of this work, the latter range is to be kept in mind as a general guideline, so as  to remain in the range of validity of the  effective Lagragian with confined hadrons. The procedure will also serve as a template to show how the mixing effects disappear from the axion-gauge couplings as the axion mass is raised.  
 
 In practice, the analysis below applies identically to  a true heavy axion which solves the strong CP problem and to a gluonic ALP, that is, any ALP 
 which has both  electroweak and gluonic anomalous couplings, even if not related to a solution to the strong CP problem.  
  For the sake of generality, consider the addition of an extra mass term to the effective Lagrangian obtained after EW symmetry breaking but above confinement in Eq.~(\ref{Eq:Lgeneral-after}), 
\begin{equation}
\mathcal{\delta L}_a= \frac{1}{2}\,M^2 \hat{a}^2.
\label{extrama}
\end{equation}
\begin{figure}[t]
\centering
\includegraphics[width = 0.6\linewidth]{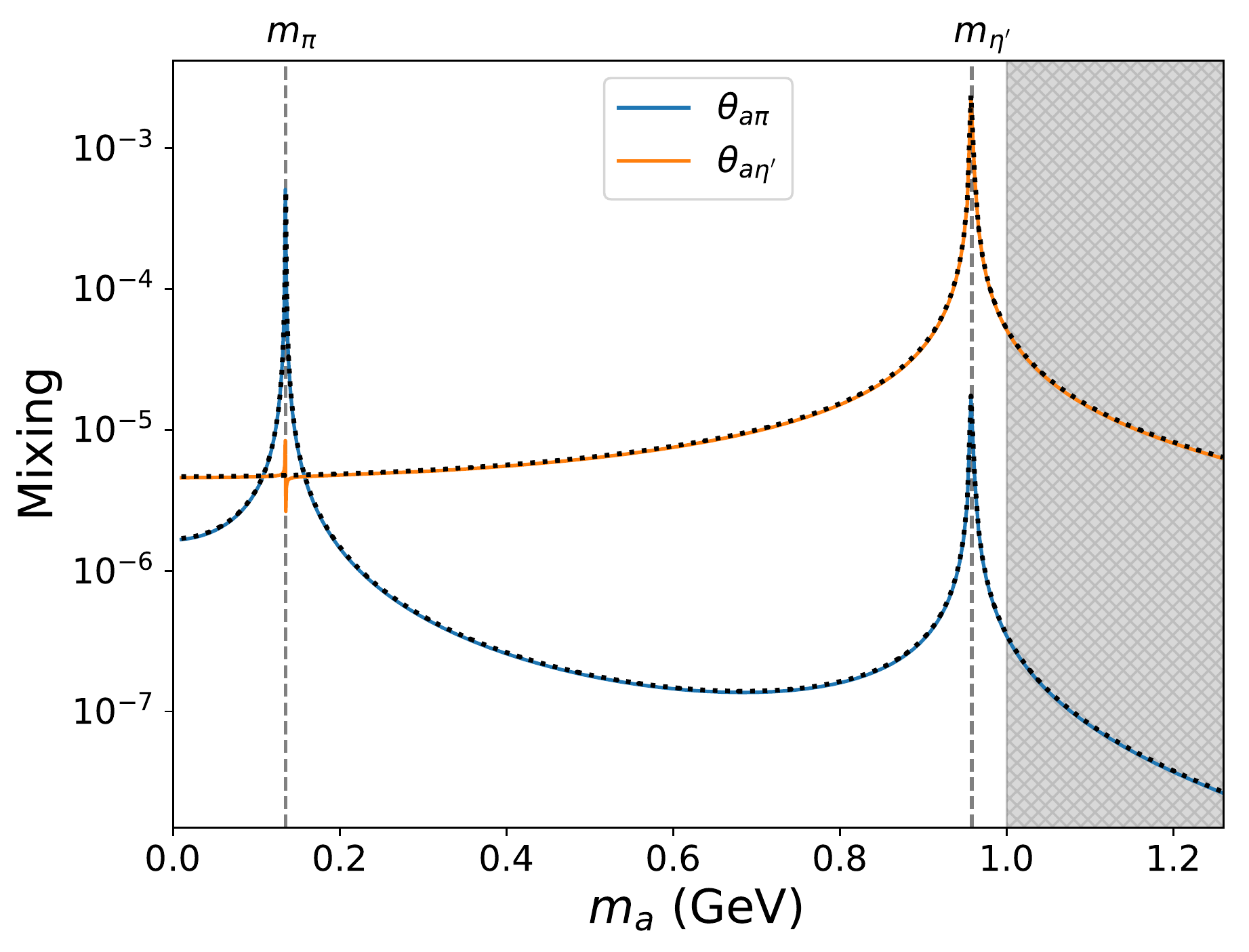}
\caption{Mixing angles as a function of the (heavy) axion mass $m_a$, for a value of $f_a=10\,$TeV. The grey area indicates the range where the validity of the chiral expansion breaks down. The figure also applies to gluonic ALPs.} 
\label{Mixings}
\end{figure}
For simplicity, from now on we focus on the case in which the first generation of SM quarks carry no PQ charge, as it is straightforward to enlarge the analysis beyond this hypothesis, as shown in the previous section. 
 After confinement, the pseudoscalar mass matrix in Eq.~(\ref{Eq:massmatrix}) is then replaced by
\beq
M^2_{\{\pi_3,\,\eta_{0},a\}}=\left(
\begin{array}{ccc}
B_0\,(m_u+m_d)  & \quad B_0\,(m_u-m_d)&0\\ 
B_0\,(m_u-m_d)  &  \quad 4K/
{f_\pi}+B_0(m_u+m_d) &\quad {2K}/({f_\pi f_a})\\ 
0 & {2 K}/({f_\pi f_a})  &  {K}/{f_a^2}+ M^2
\end{array}  
\right)\,.\label{Eq:massmatrixHeavy}
\eeq
In the limit   $f_a \gg K^{1/4}$ (i.e. $f_a \gg \Lambda_{QCD}$) the expressions for the $\pi$, $\eta'$ and $a$  mass eigenvalues become
 \begin{align}
m^2_{\pi}\simeq B_0\,(m_u+m_d),\quad m^2_{\eta'}\simeq\frac{4K}{f^2_\pi}+B_0\,(m_u+m_d), \quad  m^2_{a}\simeq M^2+ \frac{f_\pi^2m_\pi^2}{f_a^2}\frac{m_um_d}{(m_u+m_d)^2}\,.
\label{masses-heavy}
\end{align} 
The corresponding axion-pion and axion-$\eta'$  mixing angles take a very simple form in the limit $M\gg m_\pi$, 
\begin{align}
\theta_{a\pi}\simeq -\frac{f_\pi}{2f_a}\frac{m_d-m_u}{m_u+m_d}\,  \frac{1}{1-\frac{M^2}{m^2_\pi}} \, \frac{1}{1-\frac{M^2}{m^2_{\eta'}}}\,,  \qquad \theta_{a{\eta'}} \simeq -\frac{f_\pi}{2f_a}\,\frac{1}{1-\frac{M^2}{m^2_{\eta'}}}\,, \label{mixingangles-extra}
\end{align}
where again only the leading term on each entry of the mixing matrix has been retained. In fact, it can be checked that these equations hold even for small values of $M$, as long as M is non-degenerate with the pion or $\eta'$ mass.  The comparison with Eq.~(\ref{thetas-1}) ilustrates that  the M-dependent corrections may be important for axion masses near the pion mass or the $\eta'$ mass. The  divergences in Eq.~(\ref{mixingangles-extra}) for an axion degenerate in mass with the pion or the $\eta'$ are an artifact of the approximations which in practice correspond to large mixing values, as expected in those ranges.     The numerical results do not rely on that approximation and are illustrated in Fig.~\ref{Mixings}: the spikes correspond to an axion degenerate with either the pion or  the $\eta'$.

\subsection{ Heavy axion couplings to EW gauge bosons}
The couplings of the heavy axion to the electroweak gauge bosons reflect the dependence of the mixing parameters on the extra source of mass $M$, as follows: 

\begin{align}
g_{aWW} &= -\frac{1}{2\pi f_a} \frac{\alphaem}{s_w^2} 
	\bigg( \frac{L}{N} - \frac{3}{4}\frac{1}{1-\big(\frac{M}{m_{\eta'}}\big)^2} 
	\bigg)\,,\label{gaWWfinal}\\
g_{a\gamma\gamma} &= -\frac{1}{2\pi f_a} \alphaem 
	\bigg[ \frac{E}{N} - \frac{1}{1-\big(\frac{M}{m_{\eta'}}\big)^2}
		\bigg(\frac{5}{3} + \frac{m_d-m_u}{m_u+m_d} \,\frac{1}{1-\big(\frac{M}{m_{\pi}}\big)^2}
		\bigg)
	\bigg]\,,\label{gagammafinal}\\
g_{aZZ} &= -\frac{1}{2\pi f_a} \frac{\alphaem}{s_w^2c_w^2} 
	\bigg[ \frac{Z}{N} - \frac{1}{1-\big(\frac{M}{m_{\eta'}}\big)^2}
		\bigg(\frac{11s_w^4+9c_w^4}{12} - \frac{s_w^2(s_w^2-c_w^2)}{2}\frac{m_d-m_u}{m_u+m_d}  
				\frac{1}{1-\big(\frac{M}{m_{\pi}}\big)^2}
		\bigg)
	\bigg]\,,\\
g_{a\gamma Z} &= -\frac{1}{2\pi f_a} \frac{\alphaem}{s_wc_w} 
	\bigg[ \frac{2K}{N} -  \frac{1}{1-\big(\frac{M}{m_{\eta'}}\big)^2}\,
		\bigg(\frac{9c_w^2-11s_w^2}{6} - {\frac{1}{2}}(c_w^2-3s_w^2)\frac{m_d-m_u}{m_u+m_d} 
			\frac{1}{1-\big(\frac{M}{m_{\pi}}\big)^2}
		\bigg)
	\bigg]\,.\label{gagammaZfinal}	
\end{align}
The $M$-dependent corrections in these couplings can be relevant for heavy axions which solve the strong CP problem as well as for gluonic ALPs, as long as their mass parametrized by Eq.~(\ref{masses-heavy}) is sensibly larger than that for the QCD (i.e. invisible) axion, $M^2>m_\pi^2 f_\pi^2/f_a^2$. These expressions hold as long as chiral perturbation theory is valid, that is $M\lesssim 1\,$GeV.
 
 For values of $M$ noticeably larger than the $\eta'$ mass the model prediction depicted is only indicative, as the effective Lagrangian in terms of pions and $\eta'$ is not really adequate and the description should the be done in terms of the couplings to quarks. At those energies ($m_a\gg m_{\eta'}$) QCD is perturbative and  it would be pertinent  to compute the two-loop contribution of the gluonic coupling to the EW gauge boson couplings. For the case of photons, a qualitative  estimation has been performed in Ref.~\cite{Bauer:2017ris} with the result:
 \begin{equation}
  \delta g_{a\gamma\gamma} = - \frac{3\,\alpha_{\rm em}\,\alpha_s(m_a^2)}{\pi^2}\, g_{agg} \,\sum_f q_f^2 B_1(\tau_f) \,\text{log}\bigg(\frac{f_{\rm PQ}^2}{m_f^2} \bigg)  \,,
  \label{perturbative}
  \end{equation}
  leading to a photonic axion coupling given by
  \begin{equation}
  g_{a\gamma\gamma} = -\frac{1}{2\pi f_a} \alphaem \Bigg( \frac{E}{N} - \frac{3\,\alpha_s^2}{2\pi^2}\sum_f q_f^2 B_1(\tau_f) \,\text{log}\bigg(\frac{\fpq^2}{m_f^2} \bigg)  \Bigg)\,.
  \end{equation}

  The derivation of the equivalent formula for the coupling of axions to heavy EW gauge boson couplings is left for future work.  Nevertheless, the analysis presented here conveys  the qualitative behaviour expected for the transition between the low and high axion mass regimes.
 
 \subsubsection*{Gauge invariance}
For high axion  masses (i.e. $M\gg \Lambda_{QCD}$ in Eq.~(\ref{masses-heavy})), the mixing of the axion with the SM pseudoscalars becomes negligible. For those scales, QCD enters the perturbative regime and Eq.~(\ref{perturbative}) illustrates how the model-independent contributions to the EW couplings diminish. As the latter become negligible, 
  the axion coupling to EW gauge bosons is  parametrized by just the two effective interactions in Eqs.~(\ref{Wtilde}) and (\ref{ABtilde}).   
In other words, the heavy axion EW couplings   span a  parameter space with two degrees of freedom (instead of three for light axions with $m_a\ll\Lambda_{QCD}$, see Sect.~\ref{axion-EW}). Two independent constraints follow  for heavy axions: 
\begin{align}
g_{aWW}&= g_{a\gamma\gamma} + \frac{c_w}{2\,s_w}g_{a\gamma Z}\,,
\label{constrain-heavy-I}\\
g_{aZZ}&= g_{a\gamma\gamma} + \frac{c^2_w-s^2_w}{2\,s_wc_w}g_{a\gamma Z} \,, \label{constrain-heavy-II}
\end{align}
where we have chosen to express the couplings $g_{aWW},\,g_{aZZ}$ in terms of the overall better constrained ones $g_{a\gamma\gamma}$ and $g_{a\gamma Z}$.\footnote{Obviously, the milder constrain in Eq.~(\ref{constrain-light}) also applies here.}  These powerful relations will be exploited in the next section to further constrain  uncharted regions of the experimental parameter space.  

Alike to the discussion after Eq.~(\ref{constrain-light}), the relations in Eqs.~(\ref{constrain-heavy-I}) and (\ref{constrain-heavy-II})  apply not only to heavy axions and heavy gluonic ALPs,  but also to generic ALPs which only exhibit EW interactions and are much heavier than nucleons. The corollary that at least two EW couplings --if any--  must exist for any axion or ALP  holds as well.

\section{Phenomenological analysis}

The impact of the results obtained above on present and future axion searches will be illustrated in this section. Both tree-level and loop-level effects will be taken into account. Indeed the latter are relevant when confronting data on photons, electrons and nucleons, as the experimental constraints on these channels are so strong that they often dominate the bounds on EW axion couplings.

\subsection{Loop-induced couplings}\label{sec:loop_induced_couplings}
The tree-level coupling of the axion to photons can be suppressed in some situations~\cite{DiLuzio:2016sbl, DiLuzio:2017pfr} (photophobic ALPs are also possible~\cite{Craig:2018kne}).
Additionally, many models have no tree-level couplings to leptons or suppressed couplings to nucleons.~\cite{DiLuzio:2017ogq}
 However, all possible effective couplings will mix at the loop level. This affects the renormalization group (RG) evolution, via which all couplings allowed by symmetry will be generated even when assuming  only a subset of couplings at some scale.
 
Before proceeding with the phenomenological analysis, we discuss in this subsection the loop-induced effective interactions arising from the direct coupling to electroweak gauge bosons.
Because the experimental and observational limits are usually strongest for photons, electrons, and nucleons, the loop-induced contributions to these channels can give stronger constraints than those stemming from the tree-level impact on other channels.

Denoting the effective axion-fermion Lagrangian by 
\begin{equation}
\delta \LL_{a\,\rm eff} \supset \sum_{f}c_{1\,\mathrm{eff}}^f\,\frac{\de_\mu \hat{a}}{f_{\text{PQ}}} \left(\bar{f} \,\gamma_\mu\,\gamma_5\, f \,\right)\,,
\end{equation}
it has been shown~\cite{Bauer:2017ris} that the coefficient ${c_{1\,\mathrm{eff}}^f}$ receives one loop-induced corrections from electroweak couplings,
\begin{equation}\label{eq:1-loop_fermion_coupling}
\begin{aligned}
\frac{ c^f_{1\,\rm eff}}{f_a} &= \frac{3}{4}\left(\frac{\alpha_{\rm em}}{4\pi\,s_w^2} \frac{3}{4} g_{aWW} + \frac{\alpha_{\rm em}}{4\pi\,c_w^2}\left( Y_{f_L}^2 + Y_{f_R}^2 \right) g_{aBB} \right)\log{\frac{f_{\rm PQ}^2}{m_W^2}}\ +\ \frac{3}{2} \frac{\alpha_{\rm em}}{4\pi} Q_f^2 g_{a\gamma\gamma} \log{\frac{m_W^2}{m_f^2}}\\
&= \frac{9}{16} \frac{\alpha_{\rm em}}{4\pi\,s_w^2} g_{aWW} \log{\frac{f_{\rm PQ}^2}{m_W^2}}\ +\ \frac{3}{4} \frac{\alpha_{\rm em}}{4\pi\,s_w^2c_w^2}\left( \frac{3}{4}c_W^4 + \left( Y_{f_L}^2 + Y_{f_R}^2 \right)s_W^4 \right) g_{aZZ} \log{\frac{f_{\rm PQ}^2}{m_W^2}}\\
&+ \frac{3}{4} \frac{\alpha_{\rm em}}{4\pi\, s_wc_w}\left( \frac{3}{4}c_W^2 - \left( Y_{f_L}^2 + Y_{f_R}^2 \right)s_W^2 \right) g_{a\gamma Z} \log{\frac{f_{\rm PQ}^2}{m_W^2}}\ +\ \frac{3}{2} \frac{\alpha_{\rm em}}{4\pi} Q_f^2 g_{a\gamma\gamma} \log{\frac{f_{\rm PQ}^2}{m_f^2}}\,.
\end{aligned}
\end{equation}
To obtain this result, the ultraviolet scale  inside the loop diagrams has been identified with the axion scale $f_{\rm PQ}$, which is the cutoff of the effective theory. 
Note that the one-loop induced contributions to the fermion couplings are independent of the axion mass (other than a negligible  dependence through the axion-gauge couplings such as $g_{a\gamma\gamma}$, see below). They are generically of order $\alpha/4\pi$ as expected, that is, over two orders of magnitude smaller than the original effective gauge coupling. Nevertheless, the experimental constraints are so strong that they will often provide the leading bounds on gauge-axion couplings. 

The most relevant fermionic limits are those on the  coupling to electrons and light quarks.
While Eq.~\eqref{eq:1-loop_fermion_coupling} is directly applicable to leptons and heavy quarks, at low energies light quarks form hadrons:   the loop-induced coupling to nucleons have highest impact. Following Refs.~\cite{DiLuzio:2017ogq,diCortona:2015ldu}, the relation between nucleon and light quark couplings can be written as  
\begin{equation}
\begin{aligned}
c_p + c_n &= 0.50(5)\left( c^u_1 + c^d_1 -1 \right) - 2\delta\,, \\
c_p - c_n &= 1.273(2)\left( c^u_1-c^d_1 - \frac{1-z}{1+z} \right)\,,
\label{cpcn}
\end{aligned}
\end{equation}
where $z=m_u/m_d=0.48(3)$ and  $c^u_1$ and $c^d_1$ are defined in terms of the coefficients in Eq.~(\ref{general-NLOLag-c1}) as
\begin{equation}
\begin{aligned}
c^u_1 &= \frac{c_1^U  - c_1^Q}{2}\,, \qquad
c^d_1 &= \frac{c_1^D  - c_1^Q}{2}\,.
\label{c1EWafter}
\end{aligned}
\end{equation}
In Eq.~(\ref{cpcn}), $\delta$ is a combination of the heavy SM quark coeficients analogous to those in Eq.~(\ref{c1EWafter}), $\delta = 0.038(5)c_1^s + 0.012(5)c_1^c + 0.009(2)c_1^b + 0.0035(4)c_1^t$. 

The combination of Eqs.~\eqref{eq:1-loop_fermion_coupling} and (\ref{cpcn}) allows to derive the  coupling to nucleons  induced by the coupling to electroweak bosons.

\bigskip

The axion-photon coupling also receives one-loop corrections in the presence of couplings of the axion to fermions or gauge bosons. For  energies or masses in the loops higher   
than $\Lambda_{\rm QCD}$,  quarks are the appropiate propagating degrees of freedom,\footnote{{For energies below the QCD scale,  radiative corrections involving light quarks have to be evaluated using chiral Lagrangian methods. This was achieved in Ref.\cite{Bauer:2017ris}, the conclusion being that the results remain qualitatively right once the quarks masses are  substituted by an appropriate hadronic scale, $m_\pi$ for $u$ and $d$ and $m_\eta$ for $s$. However, in the presence of gluonic couplings this contribution is subdominant to the one computed in the previous sections and will thus not be considered here.}} and the one-loop contributions for an on-shell axion can be expressed as~\cite{Bauer:2017ris}
\begin{equation} \label{eq:1-loop_photon_coupling}
g_{a\gamma\gamma}^{\rm eff} = g_{a\gamma\gamma}^0 + \sum_{f}N_C^f Q_f^2\frac{\alpha_{em}}{\pi} \frac{2c_1^f}{f_a} B_1({\tau_f}) + 2 \frac{\alpha_{em}}{\pi} g_{aWW}B_2(\tau_W),
\end{equation}
where the subscript $f$ runs over leptons and heavy quarks. Here, $\tau_i = 4m_i^2/m_a^2$ and 
\begin{equation}
\begin{aligned}
B_1(\tau) &= 1-\tau f^2(\tau)\\
B_2(\tau) &= 1-(\tau-1)f^2(\tau)
\end{aligned}\quad,\qquad f(\tau) =
\begin{cases}
\arcsin{\frac{1}{\sqrt{\tau}}} &,\quad \tau\geq 1\\
\frac{\pi}{2} + \frac{i}{2}\log{\frac{1+\sqrt{1-\tau}}{1-\sqrt{1-\tau}}} &,\quad \tau<1
\end{cases}.
\end{equation}
Asymptotically, $B_1\rightarrow 1$ in the limit $m_a\gg m_f$ and $B_1\rightarrow -m_a^2/(12m_f^2)$ when $m_a\ll m_f$.
This means that the contribution of fermions heavier than the axion is strongly suppressed.
Similarly, $B_2\rightarrow 1+\pi^2/4 - \log^2m_a/m_W$ when $m_a\gg m_W$, whereas $B_2\rightarrow m_a^2/(6m_W^2)$ in the $m_a\ll m_W$ limit.

It is worth noting that chaining the two previous one-loop contributions gives an approximate estimation of the two-loop contribution of a given  heavy gauge boson coupling to the axion-photon coupling. As an example, consider $g_{aWW}$ in  Eq.~\eqref{eq:1-loop_fermion_coupling}: it results in an effective fermion coupling $c_{1\,\mathrm{eff}}^{f}$ which, when subsequently inserted in the second term in Eq.~(\ref{eq:1-loop_photon_coupling}), results in an effective axion-photon coupling.  This can be compared with the third term which gives directly a one-loop contribution of $g_{aWW}$ to $g_{a\gamma\gamma}$: 
 for $m_a\ll m_W$, the  second term in Eq.~(\ref{eq:1-loop_photon_coupling}) may  actually be numerically larger than the third term, that is, the two-loop contribution via fermionic couplings may dominate over the one-loop gauge one, as it was pointed out in Refs.~\cite{Bauer:2017ris,Craig:2018kne}. Indeed, this two-loop contribution may be phenomenologically the most relevant one to constrain the axion couplings to heavy electroweak gauge bosons.  A caveat is that only a true two-loop computation may settle the dominant pattern, but the analysis discussed is expected to provide an order of magnitude estimate. 
\begin{figure}[t]
\centering
  \includegraphics[width=0.6\linewidth]{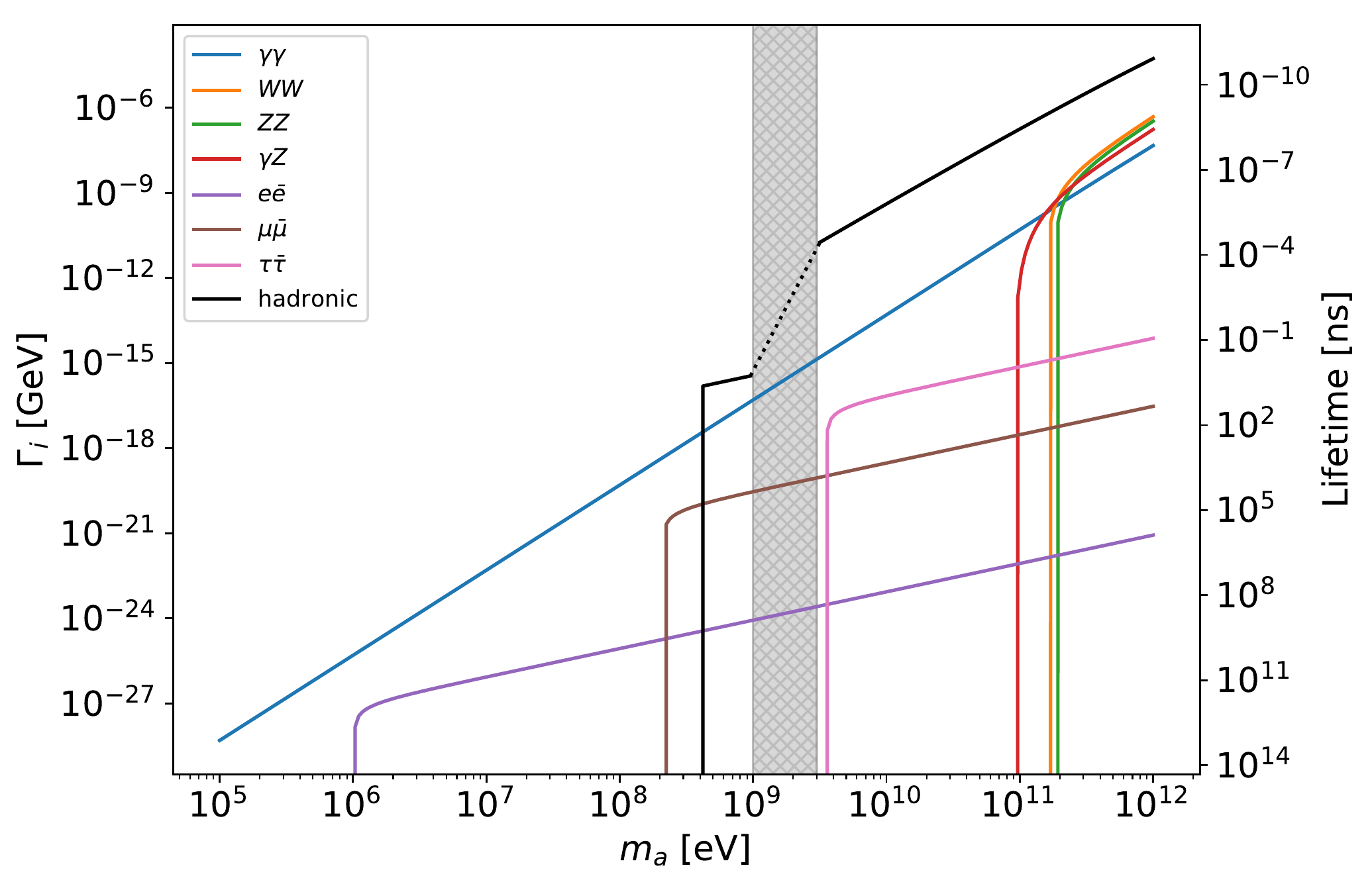}
  \caption{Illustration of the axion (or ALP) decay widths, as a function of the axion mass for a benchmark value of  $g_{a\gamma\gamma} = 10^{-7}$ GeV$^{-1}$  and the rest of gauge boson couplings as in Eq.~(\ref{Eq: relation gagg vs gaXY}).
 The grey hatched region signals the  $1$-$3$ GeV ``no human's land''  for hadronic decays in between the chiral and the perturbative regimes, where the dots indicate that no reliable prediction is possible.  }
  \label{fig:decays_qcd} 
\end{figure}

\subsection{Axion decay channels and lifetime}

The plethora of couplings discussed, contributing either at tree or loop level, produces a rich variety of production and decay channels of the axion, depending on its mass and on the relative strength of the couplings.
A quantitative evaluation  of the lifetime and branching ratios is essential for assessing what experiments or searches are more adequate to test different regions of parameter space.

In order to determine the detection capabilities of a given final state channel, an important element is whether the axion can decay into it within the detector, or escapes and contributes to an ``invisible''  channel.  For purely illustrative purposes, Fig.~\ref{fig:decays_qcd} compares the axion decay widths into different final states for a particular choice of the model-dependent parameters.\footnote{ The value for $g_{a\gamma\gamma}$  used in this figure corresponds approximately to $f_a= 10$ TeV and $E=1$ in axion models. Rescaling for other values of the couplings can be achieved by taking into account that $\Gamma_i\propto g_{aXY}^2$.} This figure serves  to indicate the mass threshold for the different channels and  is also a good indicator of the relative width of each decay channel. The determination of the areas experimentally excluded --to be developed in Sec.~\ref{expsummary}-- will not depend on the value of the effective couplings assumed in this figure, though.\footnote{The widths used in determining the coloured regions in Figs.~\ref{fig:parameter_space} induced at tree-level derive directly from the effective $g_{aXY}$ value for each point of the parameter space.}  

 In the low mass region $m_a<3m_\pi$, only decays to pairs of electrons, muons or photons are possible.
The axion typically becomes long lived enough so as to be stable at collider and flavour experiments.
 Note that this region is particularly sensitive to a possible cancellation/suppression of the photonic coupling $g_{a\gamma\gamma}$ (this happens for instance in models of axions in which the model-dependent parameter $E/N$ partially cancels the model-independent contribution, see Eqs.~(\ref{gagamma}) and (\ref{gagammafinal})). This would suppress the decay width to photons and thus enhance the branching fraction to fermions, especially close to the respective mass thresholds.

The hadronic channel plays a central role as soon as it opens. It then dominates the decay of the axion due to the large gluonic coupling.  
 The lightest possible hadronic final state is three neutral pions.
At around the GeV scale many other final states become viable, but in this region chiral perturbation theory starts to break down and we refrain from making any quantitative predictions.\footnote{After this paper was completed, Ref.~\cite{Aloni:2018vki} appeared which discusses the hadronic decays of axions in this region $1\,\mathrm{GeV}<m_a<3\,\mathrm{GeV}$.}
At high axion masses above $3\,$GeV  the inclusive decay to hadrons can be safely estimated within perturbative QCD.\footnote{Note that heavy-flavour tagging can allow to distinguish final states involving heavy quarks, but this separation will not be taken into account here.}

At much higher energies,  tree-level decays to pairs of EW gauge bosons become possible and, though subdominant with respect to the hadronic one, will play a relevant role in collider searches.

\subsection{Experimental constraints on the (heavy) axion parameter space}
\label{expsummary} 
We have reinterpreted a number of axion searches into our framework. Far from an in-depth review, this study primarily intends to point out the relative strength of the different observables in probing different flavours and parameter regions of axion and  ALP models. Interestingly and contrary to common lore, we find that some regions of parameter space can be better tested through the axion couplings to heavy gauge bosons rather than to photons.

\begin{figure}[t]
\centering
  \includegraphics[width=0.48\linewidth]{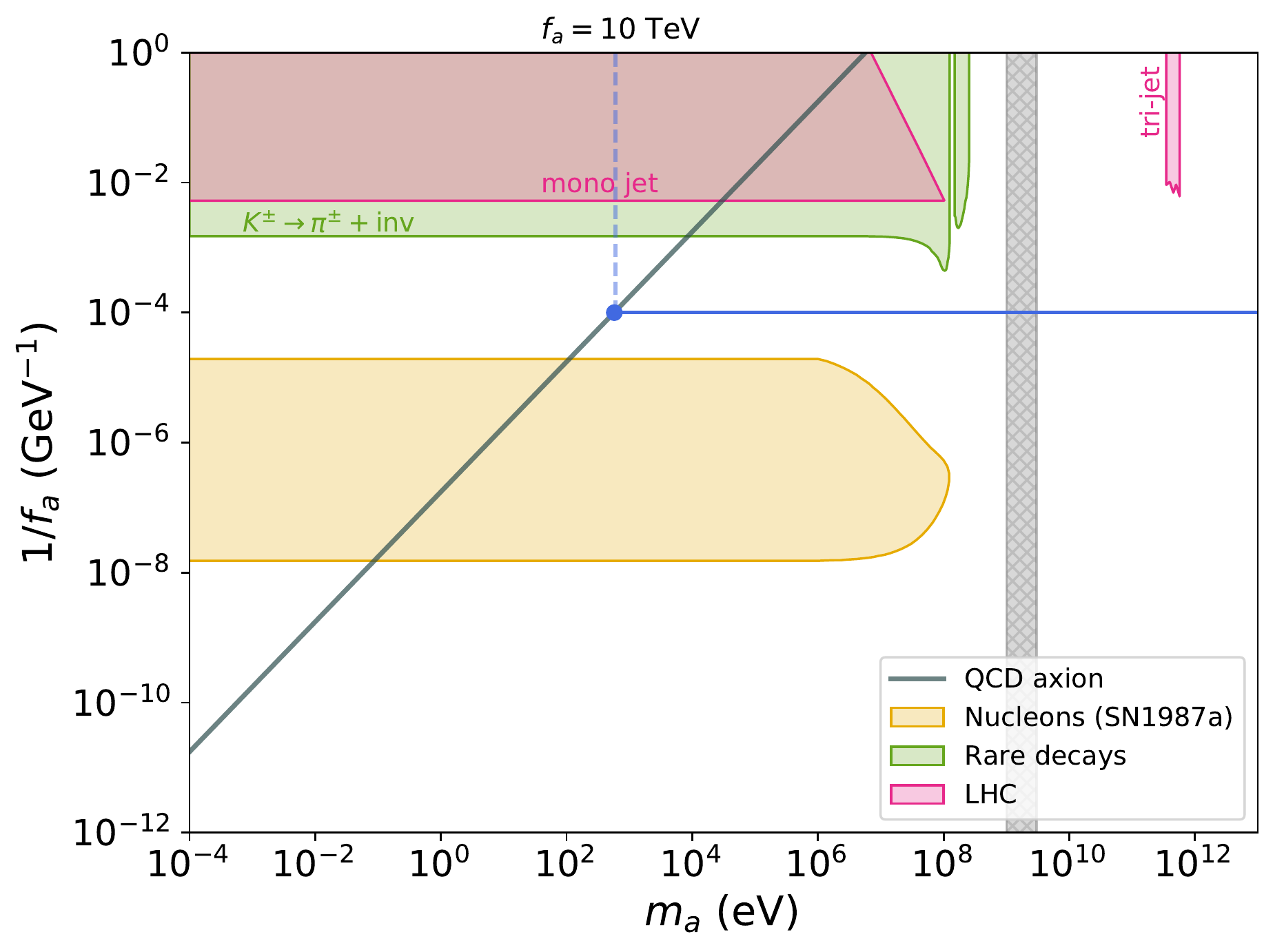}
  \caption{Coupling to gluons. The excluded areas rely on considering exclusively the coupling $g_{agg}$ by itself. The SN1987a limit was computed in Ref.~\cite{Chang:2018rso}, while the rare decays and LHC exclusions are obtained using the results of Ref.~\cite{Artamonov:2009sz} and Refs.~\cite{Mimasu:2014nea,ATLAS:2016bvn}, respectively. Particular models are represented by the overlaid lines, the black one corresponding to the QCD axion and the blue one to heavy axion models (here for a benchmark choice of $f_a=10\,$TeV). See Sec.~\ref{sec:impact_axion_models} for more details. We provide exclusion plots without superimposed lines as auxiliary files.}
  \label{fig:parameter_space_gluons} 
\end{figure} 
The coloured areas in Fig.~\ref{fig:parameter_space_gluons} show the regions experimentally excluded if taking into account {\it exclusively} the axion-gluon coupling $g_{agg}$ (which in axion models  fixes the axion scale $f_a$). Although this work focuses on the case where also EW gauge boson couplings are present, this parameter space is also shown for reference.

The couplings of axions to EW gauge bosons ($g_{a\gamma\gamma}$ $g_{aWW}$, $g_{aZZ}$ and $g_{a\gamma Z}$) will be instead explored within a two-operator framework:  {\it the axion-gluon coupling} $g_{agg}$ {\it and one electroweak gauge coupling are to be simultaneously considered}. In other words,  for each EW axion coupling  $g_{aXY}$, the regions experimentally excluded will be determined assuming the Lagrangian~\footnote{The axion mass $m_a$ is a combination of $M$ and the instanton contribution related to the first term, as previously explained (see Eq.~\ref{masses-heavy}).}
\begin{equation}
\delta\LL_a= \fourth g_{agg} \,a\,G\tilde{G} 
+ \fourth g_{aXY} \,a \,X\tilde{Y}
+\frac{1}{2}\,M^2 {a}^2\,.
\end{equation} 
 This choice is mainly motivated by the focus on solving the strong CP problem, or alternatively as an ALP analysis which goes beyond the traditional  consideration of  only one operator at a time.  The regions experimentally excluded for axion-EW gauge boson couplings are then depicted in Fig.~\ref{fig:parameter_space} as coloured areas. Note that we don't discuss any cosmological bounds. The reason for this is that models of heavy axions, typically containing an extended confining sector, are expected to significantly alter the standard cosmological picture that is usually assumed to obtain such exclusions. The study of heavy axion cosmology thus requires a full self-consistent study which is left for future work.

The resulting greenish regions in Fig.~\ref{fig:parameter_space_photons} match well-known exclusion regions for $g_{a\gamma\gamma}$, although the overlap is not complete because the latter typically only take into account the effective axion-photon coupling; the additional presence  in our analysis of the axion-gluon coupling $g_{agg}$ has a particularly relevant impact in the heavy axion region (see below).

 Figs.~\ref{fig:parameter_space_WW}, \ref{fig:parameter_space_ZZ} and \ref{fig:parameter_space_photonZ}  respectively for $g_{WW}$, $g_{ZZ}$ and $g_{\gamma Z}$ are novel.  The possibility of measuring four distinct EW observables offers a multiple window approach and a superb cross-check if a signal is detected, given the fact that 
 for axion masses much smaller (larger) than $\Lambda_{QCD}$ only three (two) couplings are independent among the set $\{$$g_{a\gamma\gamma}$ $g_{aWW}$, $g_{aZZ}$, $g_{a\gamma Z}$$\}$, see Eq.~(\ref{constrain-light})  (Eqs.~(\ref{constrain-heavy-I}) and (\ref{constrain-heavy-II})).

For the majority of the regions excluded in Fig.~\ref{fig:parameter_space}, the experiment directly constraints certain regions of the parameter space $\{g_{aXY},m_a\}$ and no further assumptions are required; those constraints apply then also to ALPs which have no gluonic couplings. However, for some collider searches the interplay between the particular EW coupling $g_{aXY}$ under analysis and the coupling to gluons plays a relevant role, and it is necessary to assume their relative strength. This will be taken as given by 
\begin{equation}
\frac{g_{agg}}{g_{aXY}}=\frac{\alpha_s}{\alpha_{XY}}\,.
\label{Eq: relation gagg vs gaXY}
\end{equation} 
This is well motivated by pseudo Nambu-Goldstone bosons with anomalous couplings generated by the triangle diagram with $\mathcal{O}(1)$ group theory factors. In any case, the results are largely insensitive to this assumption, since in the best limits stemming from LHC searches the production cross section times branching ratio scales as
\begin{equation}
\sigma(pp\rightarrow a)\times \mathrm{BR}(a\rightarrow XY) \propto \frac{g_{agg}^2\,g_{aXY}^2}{8g_{agg}^2 + g_{aXY}^2}\xrightarrow{g_{agg}\gg g_{aXY}}\frac{g_{aXY}^2}{8}\,.
\label{Eq:gagg/gaXY}
\end{equation}
It is then enough to adopt the reasonable assumption that the coupling to gluons is larger than that to EW gauge bosons $g_{agg}\gtrsim g_{aXY}$. This assumption has been adopted exclusively for LHC searches in the axion mass region where the hadronic decay channels are open, i.e. $m_a>3\,m_\pi$.

In addition, the bounds obtained via loop contributions have a logarithmic
dependence on the cut-off scale $f_a$. A relation between $f_a$ and $g_{aXY}$ needs to be assumed in order to translate the bounds on fermionic or photonic couplings into bounds on  EW gauge boson couplings. In these cases,  Eq.~(\ref{Eq: relation gagg vs gaXY}) will be used again, which for axion models translates into  
\begin{equation}
f_a=\frac{\alpha_{XY}}{2\pi \, g_{aXY}}\,,
\label{farelation}
\end{equation}
where  Eq.~(\ref{gagg}) has been used. In any case, as the cutoff dependence in loops is logarithmic this assumption has a minor impact on the exclusion plots.

To sum up,  for each EW-axion coupling the experimentally excluded regions  in Fig.~\ref{fig:parameter_space} are depicted on the parameter space $\{g_{aXY}, m_a\}$ without any assumption, except: 
\begin{enumerate}[label=\Alph*]
\item For LHC searches  and $m_a>3\,m_\pi$,  Eqs.~(\ref{Eq: relation gagg vs gaXY}) and (\ref{Eq:gagg/gaXY}) were used, which for most cases is equivalent to assume $g_{agg}\gg g_{aXY}$. 
 \label{Assumption A}
\item For the regions labelled as ``photons'', ``electrons'' and ``nucleons'' in Figs.~\ref{fig:parameter_space_WW} , \ref{fig:parameter_space_ZZ} and  \ref{fig:parameter_space_photonZ}, the loop-induced bounds have a very mild dependence on the assumption in Eq.~(\ref{farelation}). 
\label{Assumption B}
\end{enumerate}

\begin{figure}[t]{}
\centering
   \begin{subfigure}[b]{0.48\textwidth}
      \includegraphics[width=1\linewidth]{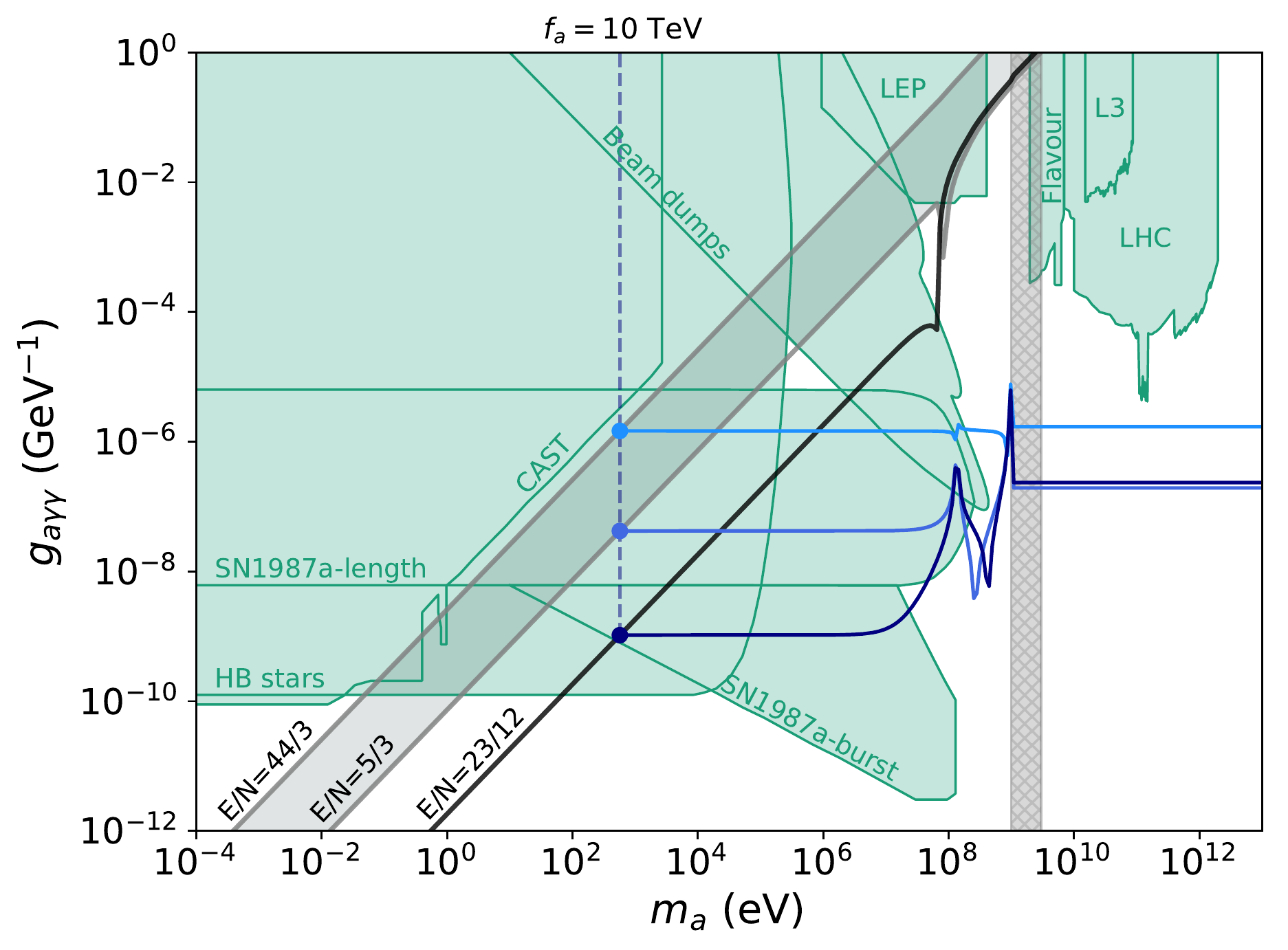}
      \caption{Coupling to photons.}
      \label{fig:parameter_space_photons} 
   \end{subfigure}
   \begin{subfigure}[b]{0.48\textwidth}
      \includegraphics[width=1\linewidth]{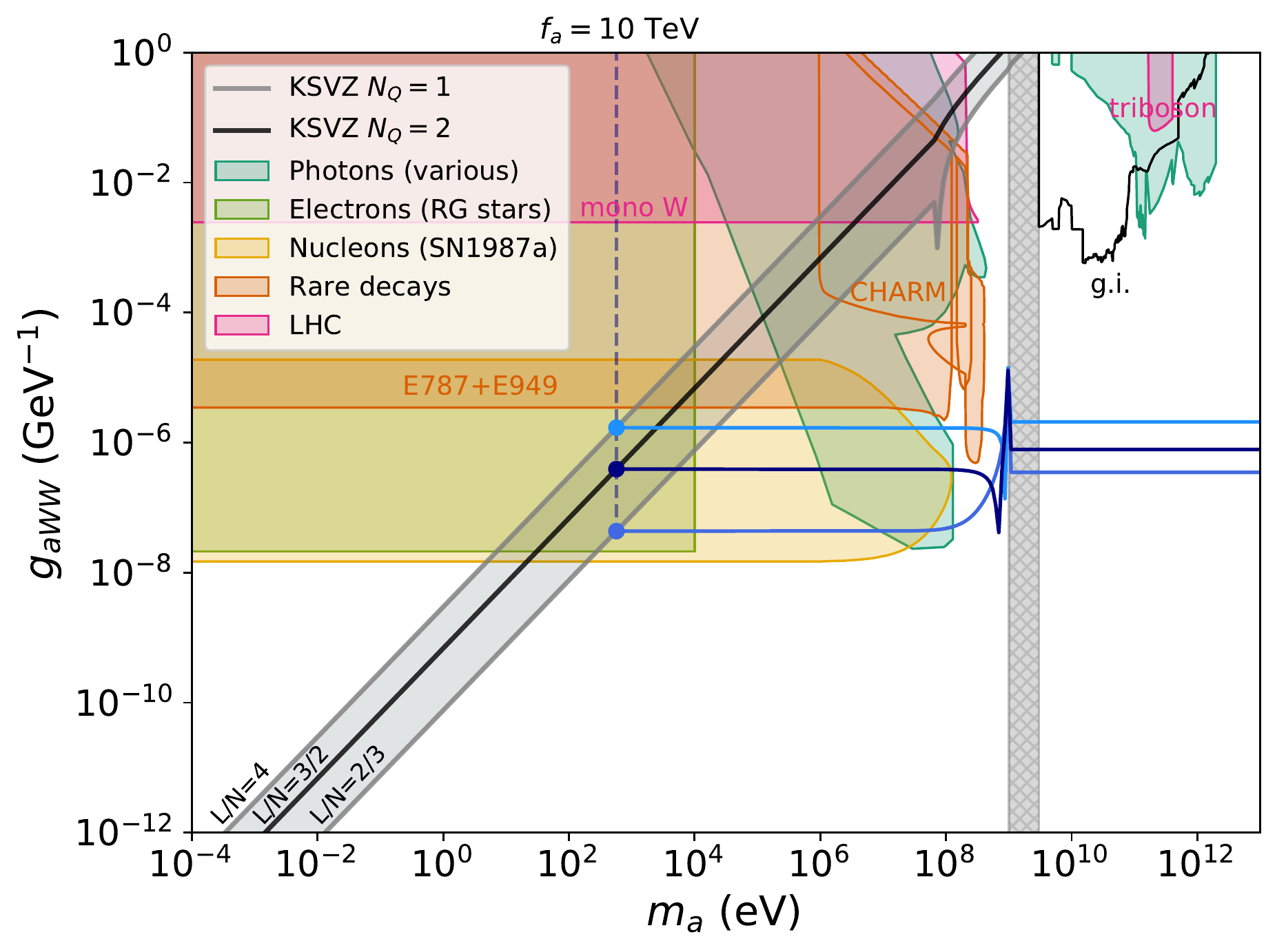}
      \caption{Coupling to $W$ bosons.}
      \label{fig:parameter_space_WW}
   \end{subfigure}

   \begin{subfigure}[b]{0.48\textwidth}
      \includegraphics[width=1\linewidth]{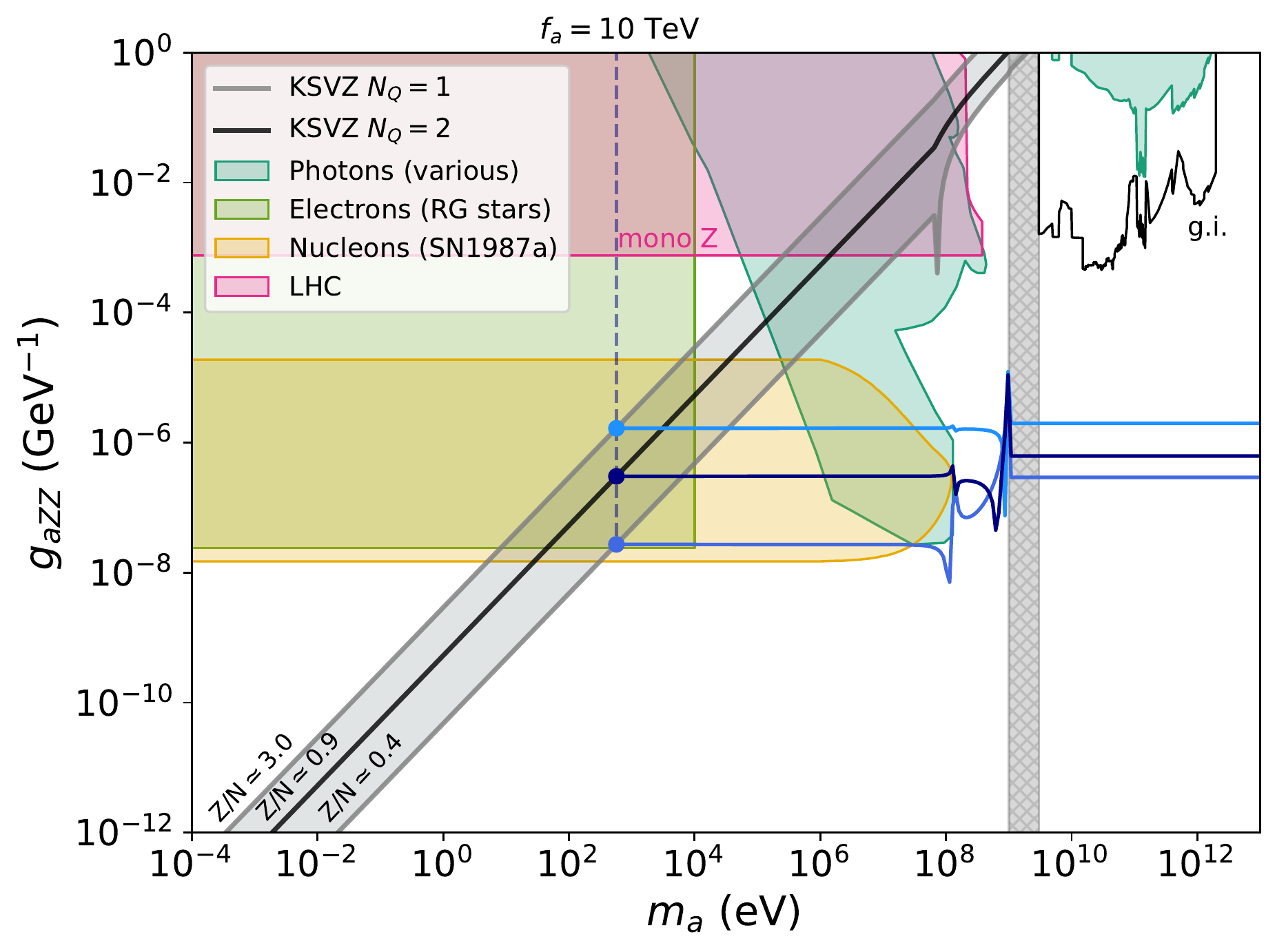}
      \caption{Coupling to $Z$ bosons.}
      \label{fig:parameter_space_ZZ}
   \end{subfigure}
   \begin{subfigure}[b]{0.48\textwidth}
      \includegraphics[width=1\linewidth]{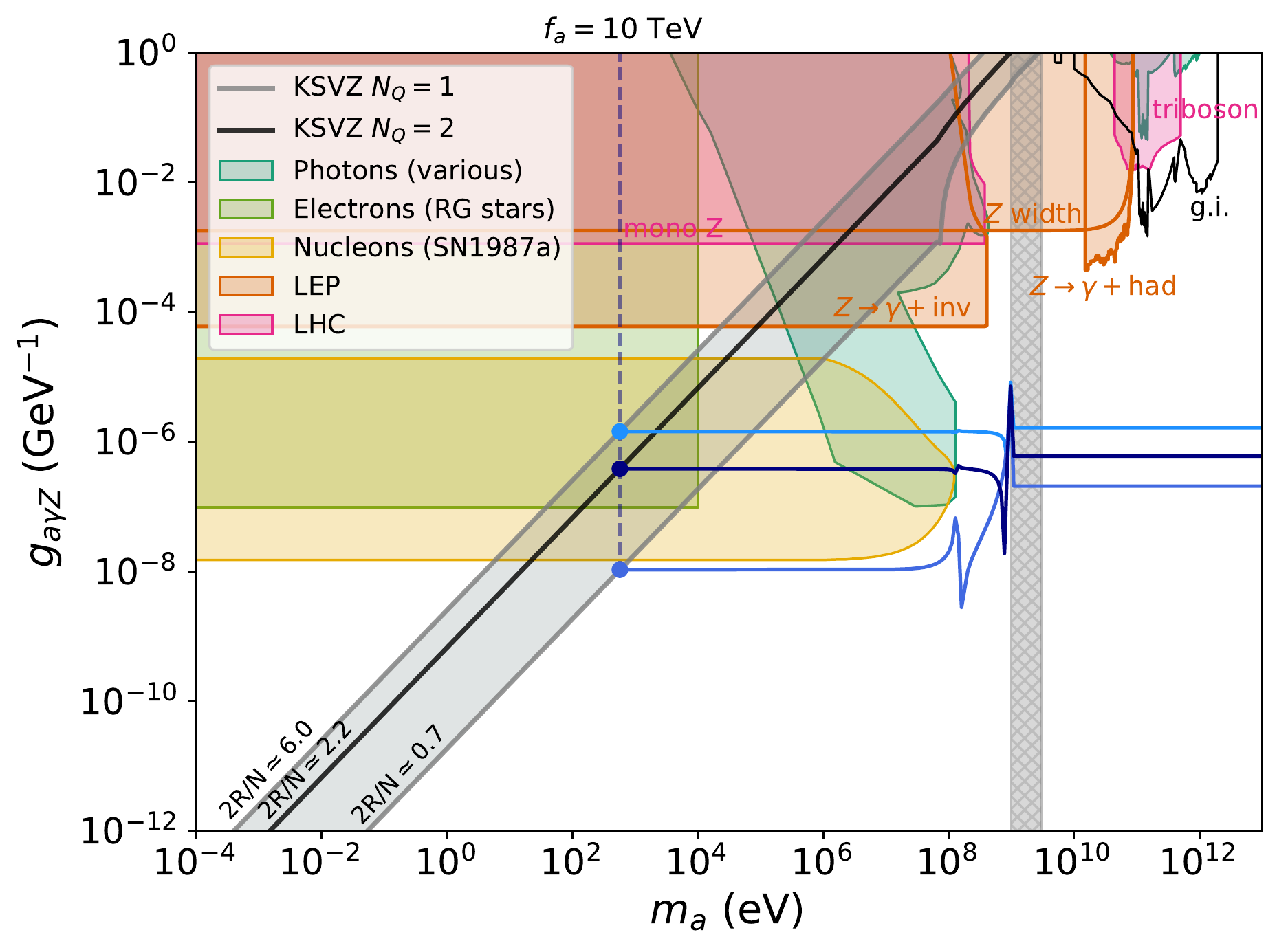}
      \caption{Coupling to a photon and a $Z$ boson.}
      \label{fig:parameter_space_photonZ}
   \end{subfigure}
\caption{ Coupling to EW gauge bosons. A two-operator framework is used:  each panel assumes 
the existence of the corresponding electroweak coupling plus the axion-gluon coupling. The coloured regions are experimentally excluded. In panels b, c and d, the loop-induced constraints labelled as ``photons'', ``electrons'' and ``nucleons'' depend mildly on $f_a$ (see text). The label ``g.i.'' illustrates the exclusion power of EW gauge invariance. The superimposed lines correspond to KSVZ-type QCD axion models   (black line and grey band) and to one benchmark heavy axion model with $f_a=10\,$TeV (blue lines). The parameter space for heavy axions corresponds to moving up and down that set of blue lines, see Sec.~\ref{sec:impact_axion_models}. The results apply as well to gluonic ALPs (we provide exclusion plots without superimposed lines as auxiliary files). }
\label{fig:parameter_space}
\end{figure}

{After having presented the general strategy that we use for the reinterpretation of constraints into our setup, we proceed to describe the origin of each of the exclusion regions coloured in Fig.~\ref{fig:parameter_space}.}

\subsubsection*{Coupling to photons}
The combination of astrophysical and terrestrial probes makes this search a particularly powerful tool to test the axion paradigm, especially for low mass axions. Even in the case of relatively large axion masses and/or situations where the coupling to photons can be suppressed, photons still place strong constraints both at tree-level and through loop-induced effects.

The experimental limits  on $g_{a\gamma\gamma}$ are compiled in Fig.~\ref{fig:parameter_space_photons}.
At the lowest axion masses considered here, $m_a \lesssim 10\rm{~meV}$, the most competitive bounds come from the CAST helioscope~\cite{Anastassopoulos:2017ftl}, and will improve in the future with the upgrade to the IAXO experiment. \cite{Armengaud:2014gea} 
At slightly larger masses up to $m_a\sim 1\,\mathrm{keV}$, $g_{a\gamma\gamma}$ is constrained by an energy-loss argument applied to Horizontal Branch (HB) stars.\cite{Raffelt:2006cw} 
A similar argument applies to the supernova SN1987a and constrains larger masses up to the $100\,$MeV range, both using extra cooling arguments~\cite{Payez:2014xsa} and by the lack of observation of a photon burst coming from decaying emitted axions.\cite{Jaeckel:2017tud} 
In the same mass range, larger couplings can be constrained using beam dump experiments, with these exclusions led at present by the NuCal~\cite{Blumlein:2013cua} experiment together with the 137~\cite{Bjorken:1988as} and 141~\cite{Riordan:1987aw} experiments at SLAC.
We adapt here the constraints compiled in Ref.~\cite{Dobrich:2015jyk}, noting that these bounds rely solely on the photon coupling.

For yet higher axion masses, where colliders provide the best limits, the gluon coupling plays a relevant role.
As long as no hadronic decay channel is open, the LEP constraints based on $Z\rightarrow\gamma\gamma$ and $Z\rightarrow\gamma\gamma\gamma$ searches obtained in~\cite{Mimasu:2014nea,Jaeckel:2015jla} for ALPs without gluonic couplings are also applicable to heavy axions.
However, for masses above $3m_\pi$, hadronic final states start to dominate and we refrain from claiming any exclusion, as a new dedicated analysis would be required to use these channels.
This explains the white gap just left of the grey hatched ``no human's land" region, which should be at least partially covered when the forementioned analysis is performed. 
It is nevertheless possible to exploit some collider searches, if a relation between the gluonic and the EW couplings is assumed. Assumption \ref{Assumption A} is adopted here.  
Our reinterpretation of the analysis in Ref.~\cite{Adriani:1992zm}, in which the $L3$ collaboration looked for hadronic final states accompanied by a hard photon, yields the limit labelled ``L3'' in Fig.~\ref{fig:parameter_space_photons}, though it is ultimately superseded by LHC exclusions.
The region labelled ``Flavour'' is excluded by data from Babar~\cite{Lees:2011wb} and LHCb~\cite{Benson:2314368}, as computed in Ref.~\cite{CidVidal:2018blh}. 
For high axion masses near the TeV scale, the limits from LHC are much stronger than those from LEP, because of the enhanced axion production via gluon-gluon fusion. We have included the limits obtained in this context in Refs.~\cite{Jaeckel:2012yz,Mariotti:2017vtv} using run 1 data.
These limits are extremely strong and should improve with the addition of run 2 data, especially at higher energies.

Finally,  the  bounds on $g_{a\gamma\gamma}$ described above have been translated --using assumption \ref{Assumption B}-- into competitive bounds for the other EW axion couplings, by means of their loop-mediated impact.

\subsubsection*{Coupling to fermions}
Flavour blind observables involving fermions can be used to constrain gauge boson couplings via the impact of the latter at loop level, see  Eq.~(\ref{eq:1-loop_fermion_coupling}). In order to fix the  mild logarithmic dependence on the cutoff scale, assumption \ref{Assumption B} will be adopted. 

The most relevant constraints on flavour-blind axion-fermion interactions are of astrophysical origin and come from either electrons or nucleons.
Firstly, a coupling of the axion to electrons allows for efficient extra cooling of some stars, which allows to place a bound on the axion-electron coupling $g_{aee}$  via the observation of Red Giants (RG)~\cite{Raffelt:2006cw}.
Secondly, and in a manner similar  to the discussion above for photons, a too strong coupling of the axion to nucleons would have shortened the duration of the neutrino burst of the supernova SN1987a.
We use the most recent evaluation of this bound calculated in Ref.~\cite{Chang:2018rso}.
These two observations (RG and SN1987a) give the strongest limits on the coupling of axions to gauge bosons for axion masses respectively below $10\,\mathrm{keV}$ and $10\,\mathrm{MeV}$,  as can be seen in Fig.~\ref{fig:parameter_space}.

In addition, the one-loop induced fermion couplings also play a role in many of the observables considered here.
In particular, they open potential axion decay channels into pairs of fermions.

\subsubsection*{Rare decays}
For axion masses in the MeV-GeV range, $g_{aWW}$ is best tested by its one-loop impact on rare meson decay experiments, where axions can be produced in flavour-changing neutral current (FCNC) processes. This search was first proposed in Ref.~\cite{Izaguirre:2016dfi} (where ALPs either stable or decaying to photons  were considered).  
{Recently, these bounds have been recomputed in Ref.~\cite{Craig:2018kne} in the context of photophobic ALPs, considering as well the potential decays of the axion to a pair of fermions due to the one-loop induced coupling in Eq.~(\ref{eq:1-loop_fermion_coupling}). We reinterpret these searches,  taking into account in addition the effects of the gluonic axion coupling under the assumptions \ref{Assumption A} and \ref{Assumption B}. The main consequence is that, for axion masses $m_a>3m_\pi$, the sensitivity is drastically reduced because of the opening of hadronic axion decay channels.} 

At low axion masses below $2m_\mu$, the axion is long-lived enough so that it can be considered stable for experimental purposes.
This means that in these regions the axion has to be looked for in invisible searches.
The most stringent limits were placed by the E787 and E949 experiments testing the $K^+\rightarrow \pi^+X$ channel, with $X$ invisible.\cite{Artamonov:2009sz} 
Following Ref.~\cite{Izaguirre:2016dfi}, we reinterpret this search in terms of axions coupled to $W$ bosons, which yields the constraint shown in  Fig.~\ref{fig:parameter_space_WW}.
These bounds will be improved in the near future by the NA62 experiment.

Axions can also be produced from rare meson decays in proton beam dump experiments, where they can be looked for in searches for long-lived particles.
The current best limits are set by the CHARM experiment~\cite{Bergsma:1985qz}, where the axion can be produced in Kaon and $B$ meson decays and  subsequently decays to a pair of electrons or muons.
The framework developed in Ref.~\cite{Clarke:2013aya} has been recast to obtain the limit on $g_{aWW}$ shown in Fig.~\ref{fig:parameter_space_WW}.\footnote{After this work was completed, Ref.~\cite{Dobrich:2018jyi} appeared which revisits the CHARM exclusion contour and provides projections of the expected NA62 and SHiP sensitivities.}

\subsubsection*{Direct couplings to heavy gauge bosons}

\paragraph{LEP} provides the best environment to directly test the $g_{a\gamma Z}$ coupling for axion masses below $m_Z$, as shown in  Fig.~\ref{fig:parameter_space_photonZ}.
The first constraint set assuming only the $g_{a\gamma Z}$ coupling was placed in Ref.~\cite{Brivio:2017ije} exploiting the limit on the uncertainty of the total $Z$ width,~\cite{ALEPH:2005ab} $\Gamma(Z\rightarrow\mathrm{BSM})\lesssim 2\,\mathrm{MeV}$ at $95\%$ C.L., which allows to set a conservative bound on the process $Z\rightarrow a\gamma$.
Stronger limits can be placed by more specific searches, as studied in Ref.~\cite{Craig:2018kne}.
The best limit at axion masses low enough for the axion to be long-lived stem from the $Z\rightarrow\gamma+\,\mathrm{inv}.$ search.
 For higher axion masses, the large hadronic branching fraction makes the $Z\rightarrow \gamma+\,\mathrm{had.}$ search the more fruitful one to look for axions.
Under assumption  \ref{Assumption A} for the relative strength of the gluonic and EW couplings, we exploit the results of the search performed by the L3 collaboration as presented in Ref.~\cite{Adriani:1992zm} to obtain strong limits for $m_a$ in the range from $10\,$GeV up to the $Z$ mass.
Note that, even if the search is the same than that used to constrain the photonic axion coupling, the exclusion for $g_{a\gamma Z}$ has a larger reach due to the fact that the process is mediated by an on-shell $Z$ boson, instead of a very virtual photon.

\paragraph{LHC} allows to look for a plethora of processes sensitive to axions. In particular, for heavy axions it provides the best limits  on the coupling to heavy EW gauge bosons. 
The drawback of restricting the analysis to processes that separately involve only one of the EW gauge boson couplings plus the gluon coupling is 
the reduced number of available searches. Nevertheless, the advantage is that it provides robust and model-independent constraints.

The authors of Ref.~\cite{Brivio:2017ije} studied the LHC phenomenology of axions that are stable on collider lengths and thus would manifest themselves as missing energy.
In particular, mono-$W$ and mono-$Z$ final states where an axion is radiated as missing energy/momentum can set constraints on the three couplings $g_{aWW}$, $g_{aZZ}$ and $g_{a\gamma Z}$, as shown in Fig.~\ref{fig:parameter_space}.
For large axion masses $m_a>m_Z$, the authors of Ref.~\cite{Craig:2018kne} suggested that triboson final states place the strongest bounds on ALPs coupling to massive gauge bosons, though the sensitivity of this search is hindered for axions because of the large hadronic branching ratio that we take into account.
Adapting their constraints --with assumption \ref{Assumption A}-- leads to the exclusion of regions in parameter space near the TeV range, as shown in Figs.~\ref{fig:parameter_space_WW} and \ref{fig:parameter_space_photonZ} for $g_{aWW}$ and $g_{a\gamma Z}$, respectively.
 Note that significant exclusions can also be placed through the loop-induced coupling to photons.
However, the most promising LHC search is one that --to the best of our knowledge-- has not been performed yet.
We advocate~\cite{Troconiz2018} the use of $pp\rightarrow a \rightarrow VV^\prime$ processes, which benefit from the large production cross section through the gluonic coupling together with the clean final states that the decay to EW gauge bosons produce.
We foresee that this search will have a sensitivity to the couplings of axions to heavy EW gauge bosons similar to the photonic case presented in Fig.~\ref{fig:parameter_space_photons}. 
Though potentially very interesting, the detailed analysis that this study requires is beyond the scope of this work and is left for the future.\cite{Troconiz2018}

\subsection{Impact on (heavy) axion models and gluonic ALPs}\label{sec:impact_axion_models}
The black oblique line in Fig.~\ref{fig:parameter_space_gluons} corresponds to the linear relation between $1/f_a$ and $m_a$ for the QCD axion, Eq.~(\ref{invisiblesaxion}). The horizontal blue branch is one example of how that relation changes after Eq.~(\ref{masses-heavy}) for an illustrative example of a true heavy axion.

The black, grey and blue lines in Fig.~\ref{fig:parameter_space} illustrate possible $\{m_a, f_a\}$ values for axions which have a gluonic coupling $g_{agg}$ (and thus may solve the strong CP problem) in addition to at least one coupling to heavy gauge bosons. Those model-dependent lines are superimposed\footnote{For the reader interested in generic gluonic ALPs rather than heavy axions, we provide the exclusion plots without any superimposed lines as auxiliary files.} on the  colored/white regions excluded/allowed by experiments for each one of the couplings in the set $\{g_{a\gamma \gamma}$, $g_{a\gamma Z}$, $g_{aZZ}, g_{aWW} \}$, as determined above. 
The examples chosen corresponds to KSVZ-type axions: either a standard QCD axion or a heavy axion  as in theories with an enlarged confining gauge sector.  
\begin{table}[h!]
\begin{align*}
\begin{array}{cc|c}
   & &SU(3)_c\times SU(2)_L\times U(1)_Y    \\
\hline
\left(E/N\right)_{\rm{max}}  =& \hspace*{-10pt}44/3   & (3,3,-4/3) \\
\left(E/N\right)_{\rm{min}}  =& \hspace*{-10pt}5/3   & (3,2,+1/6) \\
 \left(L/N\right)_{\rm{max}}  =& \hspace*{-10pt}4   & (3,3,\,\,Y\,) \\
\left(L/N\right)_{\rm{min}}  =& \hspace*{-10pt}2/3   & (8,2,-1/2) \\
 \left(Z/N\right)_{\rm{max}}  =& \hspace*{-10pt}2.9   & (3,3,-4/3) \\
\left(Z/N\right)_{\rm{min}}  =& \hspace*{-10pt}0.4    & (8,2,-1/2) \\
\left(2R/N\right)_{\rm{max}}  =& \hspace*{-10pt}5.9   & (3,3,-1/3) \\
\left(2R/N\right)_{\rm{min}} =& \hspace*{-10pt}0.7   & (8,2,-1/2) 
\end{array}
\end{align*}
\caption{Maximum and minimum values of the model-dependent coefficients for the benchmark KSVZ models with only one exotic fermion representation depicted in Fig.~\ref{fig:parameter_space}.}
\label{tab:matterContent}
\end{table}

In each panel, for a given value of the model-dependent coupling:
\begin{itemize} 
\item The expectation for the pure QCD axion is depicted by grey and black lines. The bands in Fig.~\ref{fig:parameter_space}  delimited by grey lines correspond to just one exotic KSVZ fermion representation. The values of the model-dependent parameters delimiting these benchmark bands~\cite{DiLuzio:2017pfr} are summarized in Table~\ref{tab:matterContent}. 
The black line is instead an illustrative case with two fermion representations such that the coupling to photons $g_{a\gamma\gamma} $  cancels up to theoretical uncertainties.\cite{DiLuzio:2017pfr}  
 The upward bending of the lines obeys the expected change of the prediction for axion masses larger than the $\eta'$ mass,  a regime in which the last term in the parentheses in Eqs.(\ref{gagamma})-(\ref{Eq:gagammaZ}) is absent.
\item The expectations for heavy axions are illustrated with blue lines.  
 The big dots which fall on the QCD axion lines correspond to $M=0$ in Eqs.~(\ref{extrama})-(\ref{masses-heavy}). The heavy axion trajectories start on those points and the prediction moves on each blue line towards the right as $M$ grows. As the value of the axion mass gets near the pion and the $\eta'$ masses, the prediction reflects the ``resonances'' found in the pseudoscalar mixing angles and the physical couplings to gauge bosons. For larger values of $M$ the mixing effects progressively vanish, as physically expected and reflected in Eqs.~(\ref {gaWWfinal})-(\ref{gagammaZfinal}), and the predictions stabilize again. 
 The asymptotic value of the couplings is then induced only by the model-dependent parameters ($E$, $L$, $Z$, $R$), and it is often higher than for heavy axions lighter than the pion, for which the partial cancellation between the model-dependent and model-independent mixing effects may operate. 
\end{itemize}
The figures illustrate that the $M$-dependent corrections may be relevant even for not very large $M$ values. For instance, two close values 
 of the model-dependent photon couplings $E/N$ may give a very close $g_{a\gamma\gamma} $ prediction for $M$ values above the $\eta'$ mass, while that prediction can widely differ for $M$ values smaller than the QCD scale.  This is clearly reflected by the lines corresponding to  the smaller values of $E/N$ in Fig.~\ref{fig:parameter_space_photons}.
  
 The parameter space for heavy axion models spans in fact most of the region to the right of the oblique band for the QCD axion: parallel sets of horizontal lines above and below  the blue ones depicted are possible and expected for other values of the heavy axion parameters.  For a given $f_a$, varying $M$ (that is, varying $m_a$) is tantamount to move right or left on  a horizontal blue line, while varying $f_a$ displaces up or down the set of horizontal blue lines. Finally, all these considerations for heavy axion models apply as well to gluonic ALPs, as argued earlier.

 \subsubsection*{Gauge invariance}
 For heavy axions or any type of ALP with masses $m_a\gg\Lambda_{QCD}$,  couplings to EW gauge bosons are directly tested and  gauge invariance imposes the two  relations in Eqs.~(\ref{constrain-heavy-I}) and (\ref{constrain-heavy-II}). Therefore, the combination of the experimental constrains on two of the operators in the set $\{ g_{a\gamma\gamma},\, g_{aWW},\, g_{aZZ},\, g_{a\gamma Z}\}$ translates in  model independent bounds on the other two couplings.  
 For light  masses $m_a\le \Lambda_{QCD}$,  only Eq.~(\ref{constrain-light}) applies instead.  These bounds based solely on EW gauge invariance  have been depicted by black curves on the upper right corner of Figs.~\ref{fig:parameter_space_WW}, \ref{fig:parameter_space_ZZ} and \ref{fig:parameter_space_photonZ}. They susbtantially reduce the latter parameter space, especially in the cases of $g_{aWW}$ and $g_{aZZ}$, whose current direct constraints are less powerful. They are to be taken with caution, though, since in each of the exclusion plots only one EW coupling was taken into account, while the relations deduced from gauge invariance involve several non-vanishing EW couplings. Furthermore, in a future multiparameter analysis where tree-level axion-fermion couplings are included, those relations could be corrected via one-loop effects.

 \subsection{Implications for heavy axion models}
 In existing models that solve the strong CP problem with heavy axions, often either
 \begin{equation}
 M^2 \sim \Lambda'^4/f_a^2\,, \qquad \text{or} \qquad M^2\sim m_{\pi'}^2 f_{\pi'}^2/f_a^2\,
 \end{equation}
 where $\Lambda' \gg \Lambda_{QCD}$ is a new strong confining scale, and the primed fields denote exotic ``pions'' corresponding to the exotic fermions charged under the extra confining force. Let us assume the first option as an example. In this case,\footnote{ The $m_u m_d/(m_u+m_d)^2$ factor in the QCD contribution is not shown for simplicity.} 
 \begin{equation}
 \Lambda'^4 \sim  ( m_a^2 f_a^2 - m_\pi^2 f_\pi^2)\,.
 \end{equation}
 Assume that an experiment detects an axion-photon signal and a certain value for the axion mass  which correspond to a  
  point in the white region of Fig.~\ref{fig:parameter_space_photons} and located  to the right of the QCD band.  For instance, let us consider a point located on one of the flat sections of the blue lines depicted. The interpretation in terms of a heavy axion depends on whether the axion is heavier or lighter than the pion and $\eta'$, respectively:
 \begin{itemize}
 \item $m_a\gg m_\eta'$.  The model-independent effects due to the mixings with SM pseudoscalars  have become negligible, and $g_{a\gamma\gamma}$ is a  direct measure of the product $(1/f_a) E/N$. 
 \item $m_a \ll m_\pi$. In this case the measured $g_{a\gamma\gamma}$ value is undistinguishable from that  for the QCD axion,  with the $E/N$ and $f_a$ dependence given by Eq.~(\ref{gagamma}). In other words, it would indicate either a heavy axion or a QCD axion with some degree of photophobia, as in that region they become undistinguishable. This is so at least for the lowest axion masses and/or without the help of other measurements involving heavy EW gauge bosons. 
 \end{itemize}
 In both cases, in the framework of a KSVZ model, an additional measurement of the axion coupling to heavy gauge bosons would be enough to disentangle the values of $f_a$ and $M$, that is, to determine the high scale $\Lambda'$.  The model-independent corrections determined in Eqs.~(\ref{gaWWfinal})-(\ref{gagammaZfinal}) can be essential when exploiting low-energy processes (e.g. rare decays), specially when they lead to the cancellation of a given coupling. Such cancellation in a channel in general will not apply to the couplings of the axion to other gauge bosons. Overall, the fact that only two or three axion-EW couplings are independent, while four channels can be explored, will allow to overconstrain the system.

\section{Conclusions}

Among the novel results of this work, we have first determined at leading order in the chiral expansion the model-independent components of the coupling of the QCD axion to heavy EW gauge bosons:  $g_{a\gamma Z}$, $g_{aZZ}$ and $g_{aWW}$. They stem from the axion-$\eta'$-pion mixing induced by the anomalous QCD couplings of these three pseudoscalars. Our results extend to heavy EW gauge bosons the well known result for the photonic coupling of the axion $g_{a\gamma \gamma}$. They must be taken into account  whenever an axion lighter than $\Lambda_{QCD}$ is on-shell and/or the energy and momenta involved in a physical process are of the order of the QCD confining scale or lower.  As a previous step, we  re-derived pedagogically the leading contributions to $g_{a\gamma\gamma}$ for the case of the most general axion couplings (App.~\ref{App: general matrix}), and then proceeded to the determination of the couplings to the SM heavy gauge bosons. 

This analysis of the EW couplings of the QCD axion may have rich consequences when comparing the presence/absence of signals at two different energy regimes.  For instance, the axion could be photophobic at low energies~\cite{DiLuzio:2017pfr} or even EW-phobic (e.g. in rare meson decays) because of cancellations between the model-independent and model-dependent components, while an axion signal may appear at accelerators or other experiments at higher energies  at which the model-independent component disappears. 

We have next extended those results to the case of heavy axions which solve the strong CP problem. This has allowed to explore how the mixing of the axion with the pion and $\eta'$ evolves with rising axion mass, and in consequence how the model-independent contributions to all four EW axion couplings vanish as the axion mass increases above the QCD confinement scale.  We have determined the modified expression for $g_{a\gamma\gamma}$ relevant for heavy axions, which may have rich consequences: an hypothetical measurement of that coupling outside the QCD axion band could point to either a heavy axion or a photophobic QCD axion.  
The analogous expressions for  $g_{a\gamma Z}$, $g_{aZZ}$ and $g_{aWW}$  have been also worked out.

On the purely phenomenological analysis, we developed a ``two simultaneous coupling'' approach  in order to determine the regions experimentally excluded by present data for  $g_{a\gamma \gamma}$, $g_{a\gamma Z}$, $g_{aZZ}$ and $g_{aWW}$ versus the axion mass.   Each EW coupling  has been considered simultaneously  with the anomalous gluonic coupling essential to solve the strong CP problem.  The allowed/excluded experimental areas have been  depicted for each of those couplings as a function of the axion mass. This is the first such reinterpretation for $g_{a\gamma Z}$, $g_{aZZ}$ and $g_{aWW}$. 
 Even for $g_{a\gamma\gamma}$, the results of previous studies often did not apply and must be reanalysed: for instance the present bounds extracted from LEP and LHC data tend to focus on ALPs which would not have gluonic couplings, with very few exceptions~\cite{Mimasu:2014nea,Jaeckel:2012yz,Mariotti:2017vtv,CidVidal:2018blh}.  Furthermore, we have included an estimation of the one-loop induced bounds for each EW axion coupling, which leads to  supplementary constraints.

The expectations from KSVZ-type of theories have been then projected and illustrated over the obtained experimental regions, both  for the QCD axion and for heavy axions. The compatibility of hypothetical {\it a priori} contradictory signals in high and low energy experiments in terms of a given axion has been pointed out. Furthermore, we discussed how to 
 interpret an eventual signal (or null result) outside the QCD axion band  in terms of the value of the new high confining scale generically present in heavy axion theories. 

A simple point with far reaching consequences results from EW gauge invariance.  In all generality, not all couplings of axions to EW gauge bosons are independent among the four physical ones in the set  $\{g_{a\gamma\gamma}$ $g_{a\gamma Z}$, $g_{aZZ}$ and $g_{aWW}\}$. In particular, the relations obtained imply that at least two EW gauge couplings --if any--  must be non-vanishing for any axion or ALP.  These facts have been used to project the exclusion limits for the presently best constrained couplings  onto the parameter space for the less constrained ones. In particular, for axions/ALPs much heavier than $\Lambda_{QCD}$, those relations have been projected on the parameter space for $g_{aWW}$, $g_{aZZ}$  and $g_{a\gamma Z}$, reinforcing their constraints. A future multiparameter analysis may correct them via loop corrections. Nevertheless, the results obtained here clear up the uncharted experimental regions and may be of use in setting a search strategy. More in general, the existence of four physical axion couplings to EW bosons at experimental reach constitutes a phenomenal tool to over-constrain the axion parameter space and to check the origin of an eventual axion signal. 

Finally, all results obtained for heavy axions apply as well to ALPs which have both EW and gluonic anomalous couplings. The constraints stemming from EW gauge invariance extend even to generic ALPs which do not couple to gluons. In consequence, this work also extends automatically the usual parameter space for ALPs that do not intend to solve the strong CP problem, adding to the incipient efforts to go towards a multi-parameter strategy.

\section*{Acknowledgments}

We acknowledge J. Fernandez de Troconiz, R. Houtz, J. Jaeckel, J.M. No, R. del Rey, O. Sumensari and V. Sanz for very interesting conversations and comments.
We also thank J. Jaeckel for his useful remarks on the manuscript.
M.B.G and P. Q. acknowledge IPMU at Tokyo University, where the last part of this work was developed.
G. A. thanks the IFT at the Universidad Aut\'onoma de Madrid and the University of Washington for the hospitality during the first and last stages of this work, respectively.
This project has received support from the European Union's Horizon 2020 research and innovation programme under the Marie Sklodowska-Curie grant agreements No 690575  (RISE InvisiblesPlus) and  No 674896 (ITN ELUSIVES). M.B.G and P. Q. also acknowledge support from the 
 the Spanish Research Agency (Agencia Estatal de Investigaci\'on) through the grant IFT Centro de Excelencia Severo Ochoa SEV-2016-0597, as well as  from the "Spanish Agencia Estatal de Investigaci\'on" (AEI) and the EU ``Fondo Europeo de Desarrollo Regional'' (FEDER) through the project FPA2016-78645-P. 
  The work of G. A. was supported through an ESR contract of the H2020 ITN Elusives (H2020-MSCA-ITN-2015//674896-ELUSIVES) and  through a ``La Caixa'' predoctoral grant of Fundaci\'on La Caixa. The work of P.Q. was supported through a ``La Caixa-Severo Ochoa'' predoctoral grant of Fundaci\'on La Caixa.

\appendix
\newpage
\section{The neutral pseudoscalar mass matrix}
\label{masscomplete}
In the expansion in which all terms appearing at first order in $1/f_a$ and in quark masses are kept, Eq.(\ref{Eq:massmatrix}) becomes
\begin{align}
\left(\begin{matrix}  1 & 
	\frac{f \left(m_{d} - m_{u}\right)}{2 f_{a} \left(m_{d} + m_{u}\right)} \bigg(1+\frac{ m_\pi^{2}}{m^2_{\eta'}}\frac{(m_{d}- m_{u})^2 }{  \left(m_{d} + m_{u}\right)^{2}} \bigg)    & 
	\frac{f}{2 f_{a}}\bigg(1- \frac{ m_\pi^{2}}{m^2_{\eta'}}\bigg) \\ 
-\frac{f \left(m_{d} - m_{u}\right)}{2 f_{a} \left(m_{d} + m_{u}\right)}   \bigg(1-\frac{ m_\pi^{2}}{m^2_{\eta'}}\frac{4\,m_{d} m_{u} }{ \left(m_{d} + m_{u}\right)^{2}} \bigg)  &	
	 1 & 
	 - \frac{ m_\pi^{2}}{m^2_{\eta'}}\frac{ \left(m_{d} - m_{u}\right)}{\left(m_{d} + m_{u}\right)} \\ 
- \frac{f}{2 f_{a}} \bigg(1-\frac{ m_\pi^{2}}{m^2_{\eta'}}\frac{ 4\,m_{d} m_{u} }{  \left(m_{d} + m_{u}\right)^{2}} \bigg) & \frac{ m_\pi^{2}}{m^2_{\eta'}}\frac{ \left(m_{d} - m_{u}\right)}{ \left(m_{d} + m_{u}\right)} & 1\end{matrix}\right)\,.
\end{align}

\newpage

\section{Anomalous couplings of the pseudoscalar mesons to the EW gauge bosons}
\label{appendix-anomalous}

In addition to the axion couplings in Eqs.~(\ref{AGtilde})-(\ref{ABtilde}), new couplings involving gauge bosons are expected as soon as the extra pseudoscalar mesons appear in the spectrum  below the confinement scale. In particular, for the  $\eta_0$ and  the pions, the following interactions with  
 EW gauge bosons (considered as external currents) are possible below the QCD confinement scale:
\begin{eqnarray}\label{Gtilde-confined}
&\,&\WWd^a\tilde{W}^{a\mu\nu}\dfrac{{\eta_0}}{f_\pi}\,, \qquad \BBd\tilde{B}^{\mu\nu}\dfrac{{\eta_0}}{f_\pi}\,, \qquad \BBd\tilde{B}^{\mu\nu}\dfrac{{\pi_3}}{f_{\pi}}\,,\qquad \WWd^a\tilde{B}^{\mu\nu}\dfrac{{\pi_a}}{f_{\pi}}\,.\label{BW}
\label{ABtilde-confined}
\end{eqnarray}
The lightness of pseudoscalar mesons in QCD appears as a natural consequence of the spontaneous breaking of the chiral symmetry $U(2)_L\times U(2)_R\longrightarrow U(2)_V$. The pions and eta mesons can be identified as the pseudo Nambu-Goldstone bosons of the broken symmetry. However, the chiral symmetry is only approximate. It is explicitly broken not only by the quark masses, but also by the electroweak gauge interactions. This fact famously explains the difference between charged and neutral pion masses but also allows the computation of the coupling of pseudoscalar mesons to the EW bosons through the anomaly, 
\begin{equation}
\partial^\mu j^a_ \mu=\frac{\alpha_i}{8\pi}C^{abc}_{group} F_{b\,\mu\nu} \tilde F_c^{\mu\nu}\,,
 \end{equation}
where $\tilde F^{\mu\nu}=\frac{1}{2}\epsilon^{\mu\nu\sigma\rho}F_{\sigma\rho}$,  the fine structure constant of the corresponding gauge interaction is denoted by  $\alpha_i=\frac{g_i^2}{4\pi}$ and the group theoretical factor $C_{group}$ is given by
\begin{equation}
C^{abc}_{group}=\sum_{LH-RH}Tr\left[T^a\{t^b,t^c\}\right]\label{general}.
 \end{equation}
Here, $T^a$ is the generator associated to the conserved current (i.e. to the global symmetry) and $t^a$ are the generators of the  representation of the gauge group under which each fermion transforms.
 
Applying these formulas, the anomalous couplings of the neutral pseudoscalar mesons $\pi^3$ and $\eta'$ to the EW gauge bosons can be computed. For the pion, 
\begin{align}
\mathcal{L} \supset  \fourth g_{\pi BB} \,\pi_3  B\tilde{B} +\fourth g_{\pi  BW} \,\pi_3  B\tilde{W^3} \longrightarrow  \fourth g_{\pi ZZ} \,\pi_3  Z\tilde{Z} + \fourth g_{\pi \gamma\gamma} \,\pi_3  F\tilde{F} + \fourth g_{\pi \gamma Z} \,\pi_3  F\tilde{Z},
\end{align}
with
\begin{align}
&g_{\pi BB} \equiv -\frac{\alpha}{2\pi\,f_\pi} \bigg(\frac{1}{c_w^2}\bigg), 
&g_{\pi ZZ} &\equiv -\frac{\alpha}{2\pi\,f_\pi} \bigg(\frac{s_w^2}{c_w^2}-1\bigg),\nonumber\\
&g_{\pi BW} \equiv -\frac{\alpha}{2\pi\,f_\pi} \bigg(\frac{1}{c_w\,s_w}\bigg), 
&g_{\pi \gamma\gamma} &\equiv -\frac{\alpha}{\pi} \frac{1}{f_\pi}, \nonumber\\
&
&g_{\pi \gamma Z} &\equiv -\frac{\alpha}{2\pi\,f_\pi} \bigg(\frac{c_w^2-3s_w^2}{c_w\,s_w}\bigg)\, .\nonumber
\end{align}
Equivalently for the $\eta$ meson,
\begin{align}
\mathcal{L} \supset  \fourth g_{\eta'  WW} \,\eta_0\,  W\tilde{W} + \fourth g_{\eta' BB} \, \eta_0\,  B\tilde{B} \longrightarrow  \fourth g_{\eta' ZZ} \,\eta_0\,  Z\tilde{Z} + \fourth g_{\eta' \gamma\gamma} \,\eta_0\,  F\tilde{F} + \fourth g_{\eta' \gamma Z} \,\eta_0\,  F\tilde{Z},
\end{align}
with
\begin{align}
&g_{\eta' WW} \equiv -\frac{\alpha}{2\pi\,f_\pi} \bigg(\frac{3}{2}\frac{1}{s^2_w}\bigg), 
&g_{\eta' WW} &\equiv -\frac{\alpha}{2\pi\,f_\pi} \bigg(\frac{3}{2}\frac{1}{s^2_w}\bigg),\nonumber\\
&g_{\eta' BB} \equiv -\frac{\alpha}{2\pi\,f_\pi} \bigg(\frac{11}{6}\frac{1}{c_w^2}\bigg), 
& g_{\eta' ZZ} &\equiv -\frac{\alpha}{2\pi\,f_\pi} \bigg(\frac{11\,s_w^4+9\,c_w^4}{6\,s_w^2c_w^2}\bigg)\nonumber,\\
&
&g_{\eta' \gamma\gamma} &\equiv -\frac{\alpha}{2\pi\,f_\pi}\bigg(\frac{10}{3}\bigg) ,\nonumber\\
&
&g_{\eta' \gamma Z} &\equiv -\frac{\alpha}{2\pi\,f_\pi} \bigg(\frac{9\,c_w^4-11\,s_w^4}{3\,s_w^2c_w^2}\bigg)\, .\nonumber\\
\end{align}

The mixing of the axion with the $\eta'$ and the neutral pion  will result  in additional contributions to the coefficients of the interactions in Eqs.~(\ref{Wtilde}) and (\ref{ABtilde}), via the first three operators in Eq.~\ref{Gtilde-confined}. The fourth one results in a neutral pion-$W^3B$ coupling, which in turn induces a $SU(2)_L$-breaking coupling of the physical axion to $W^3B$  .
 In other words, below $\Lambda_{\rm QCD}$ the operator space of axion electroweak couplings spans three degrees of freedom, instead of the two above the QCD confinement scale, with
\begin{align}
 \delta\LL_a^{gauge}\,= &\frac{\alpha_s}{8\pi} \,\frac{a}{f_a}G\tilde{G} + \fourth g_{aWW} \,a \,W\tilde{W} + \fourth g_{aBB} \,a\, B\tilde{B} + \fourth g_{aBW} \,a\, B\tilde{W} \,, 
 \label{LunderConf-1}
\end{align}
where
\begin{equation}
\begin{aligned}
g_{aWW} &= -\frac{1}{2\pi f_a} \frac{\alphaem}{s_w^2} \left( \frac{L}{N} - \frac{3}{4} \right),\\
g_{aBB} &= -\frac{1}{2\pi f_a} \frac{\alphaem}{c_w^2} \left( \frac{P}{N} - \frac{5m_u+17m_d}{12(m_u+m_d)} \right),\\
g_{aBW} &= \frac{1}{2\pi f_a} \frac{\alphaem}{s_wc_w} \left( \frac{1}{2}\frac{m_d-m_u}{m_u+m_d} \right)\,
\end{aligned}
\end{equation}
are obtained using the result for the mass mixing of the axion with the pseudoscalar mesons given in Eq.~(\ref{aphys}). 

Similarly for the heavy axion case, 
\begin{equation}
\begin{aligned}
g_{aWW} &= -\frac{1}{2\pi f_a} \frac{\alphaem}{s_w^2} \bigg( \frac{L}{N} - \frac{3}{4}\frac{1}{1-\big(\frac{M}{m_{\eta'}}\big)^2} \bigg),\\
g_{aBB} &= -\frac{1}{2\pi f_a} \frac{\alphaem}{c_w^2} 
   \bigg[ \frac{P}{N} -\frac{1}{1-\big(\frac{M}{m_{\eta'}}\big)^2}  
       \bigg(\frac{11}{6} + \frac{1}{2} \frac{m_d-m_u}{m_u+m_d} 
          \frac{1}{1-\big(\frac{M}{m_{\pi}}\big)^2} 
       \bigg) 
   \bigg]\,,\\
g_{aBW} &= \frac{1}{2\pi f_a} \frac{\alphaem}{s_wc_w} 
  \bigg( \frac{1}{2}\frac{m_d-m_u}{m_u+m_d}
       \frac{1}{1-\big(\frac{M}{m_{\pi}}\big)^2}\,
       \frac{1}{1-\big(\frac{M}{m_{\eta'}}\big)^2}  
     \bigg)\,.
\end{aligned}
\end{equation}
\newpage

\section{How general is the mass matrix in Eq.~\ref{Eq:massmatrix}?}
\label{App: general matrix}
Let us start with the most general Lagrangian above the QCD confinement scale that is relevant for the mass and mixings of the axion:
\begin{align}
\delta\LL_a\,= &\frac{1}{2}\,\dmu \hat a\,\dmu \hat a\,+
\,c_1^u\,\frac{\de_\mu \hat{a}}{f_{\text{PQ}}} \left(\bar{u} \,\gamma_\mu\,\gamma_5\, u \,\right) \,+
\,c_1^d\, \frac{\de_\mu \hat{a}}{f_{\text{PQ}}} \left(\bar{d} \,\gamma_\mu\,\gamma_5\, d\right)\nn \\
 -&\,\bar{u}_L\, m_u\, u_R \,e^{i\,c_2^u\,\hat a/f_{\text{PQ}}} \, -
 \, \bar{d}_L\, m_d\, d_R \,e^{i\,c_2^d\,\hat a/f_{\text{PQ}}}\, +\, \hc \nn\\
 + &c_3\,\frac{\alpha_s}{8\pi} \,\frac{\hat{a}}{f_{\text{PQ}}}G\tilde{G} +  \fourth g_{a\gamma\gamma}^0 \,\hat{a}\, F\tilde{F} \,.
 \label{Eq:Lgeneral}
\end{align}
For simplicity, here we focus on the coupling to photons but the discussion is of course applicable to the other EW gauge bosons. The relation of the $\,c_1^{u,d}$ couplings with the corresponding ones $c_{1}^{Q}$, $c_{1}^{U}$ and $c_{1}^{D}$ in the $SU(2)\times U(1)$ gauge invariant formulation in Eq.~(\ref{general-NLOLag-c1}) are given by Eq.~(\ref{c1EWafter}), while $c_3=N_0$ in that equation. Without loss of generality, PQ invariance can be imposed on all operators in the Lagrangian but the anomalous couplings. As a consequence, the couplings $c_2^{u,\,d}$ are related to the PQ charges of the up and down quarks in in Eq.~(\ref{general-NLOLag-c1}) in the following way, $\mathcal{X}_{u,d}=c_2^{u,\,d}\,\mathcal{X}_a$ (where the charge of the axion can be set to $\mathcal{X}_a=1$ ).  

This Lagrangian has a reparametrization invariance \cite{Kim10p557},\footnote{Note that this reparametrization invariance differs from that in Ref.~\cite{Kim10p557}.} that corresponds to making the usual axion dependent quark field rotations,
\begin{align}
c_1^{u,d}&\to\,c_1^{u,d}+ \alpha_{u,d}\,,\nn\\
c_2^{u,d}&\to\,c_2^{u,d} - 2\,\alpha_{u,d}\,,\nn\\
c_3\,\,&\to\,c_3 + 2\, \alpha_u+2\,\alpha_d\,,\nn\\
g_{a\gamma\gamma}\,\,&\to\,g_{a\gamma\gamma} -\frac{\alpha}{2\pi \fpq}\,\left(12\alpha_u\,q_u^2+12\alpha_d\,q_d^2\right)\,.
\label{Eq:reparametrization}
\end{align}
In the body of the paper we are considering $c_1^{u,d}=0$, when computing the axion mass and mixings. This is general due to the reparametrization invariance.

When QCD confines and the chiral symmetry is broken by the quark condensate, the  couplings in Eq.~(\ref{Eq:Lgeneral}) translate into effective operators involving mesons. Defining the $\pi$ and $\eta_0$-fields in terms of the currents:
\begin{align}
j^\mu_3=\frac{1}{2}\,\left(\bar{u} \,\gamma_\mu\, \gamma_5\,u\,-\,\bar{d} \,\gamma_\mu\, \gamma_5\,d\right)\equiv f_\pi \dmu \pi_3\,,\nn \\
j^\mu_0=\frac{1}{2}\,\left(\bar{u} \,\gamma_\mu\, \gamma_5\,u\,+\,\bar{d} \,\gamma_\mu\, \gamma_5\,d\right)\equiv f_\pi \dmu \eta_0\,.
\label{Eq:pion and era currents}
\end{align}
The low energy chiral Lagrangian can be decomposed in three terms:
\begin{equation}
\delta\LL_{a}=\delta\LL_{a,\,\text{kin}}+\delta\LL_{a,\,\text{mass}}+\delta\LL_{a,\,\text{anom}}\,.
\end{equation}
\begin{itemize}
\item The derivative couplings $c_1^{u,d}$ generate kinetic mixing between the axion and the mesons,
\begin{multline}
\delta\LL_{a,\,\text{kin}}\,= \frac{1}{2}\,\dmu\hat a\,\dmup\hat a\, + 
\, \frac{1}{2}\,\dmu\pi_3\,\dmup\pi_3\, +
\, \frac{1}{2}\,\dmu\eta_0\,\dmup\eta_0\, \\
+ \,\big(c_1^u-c_1^d\big)\,\frac{f_\pi}{f_{\text{PQ}}}\,\dmu\hat a\, \dmup\pi_3  \,+
\,\big(c_1^u+c_1^d\big)\,\frac{f_\pi}{f_{\text{PQ}}}\,\dmu\hat a\, \dmup\eta_0 \,. 
 \label{Eq:LKinetic}
\end{multline}
\item The Yukawa couplings $c_2^{u,d}$  produce mass mixing. In the chiral formulation the effects of these operators can be encoded in an axion-dependent mass matrix,
\begin{align}
M_a=\left(
\begin{matrix}
m_u & 0\\
0 & m_d 
\end{matrix} 
\right)
\left(
\begin{matrix}
e^{i\,c_2^u\,\hat a/f_{\text{PQ}}} & 0\\
0 & e^{i\,c_2^d\,\hat a/f_{\text{PQ}}}
\end{matrix} 
\right)\,.
\end{align}
So that the chiral Lagrangian induced at low energies contains the term
\begin{align}
\delta\LL_{a,\,\text{mass}}\,= &\,B_0 \frac{f_{\pi}^2}{2}\, \rm{Tr}\left( \Sigma M_a^\dagger + M_a\Sigma^\dagger \right)  \nn \\
=&B_0 f_{\pi}^2 \left[
m_u\,\text{cos}\left(\frac{\pi_3}{f_\pi}+\frac{\eta_0}{f_\pi}-\,c_2^u\,\frac{\hat a}{\fpq}\right)+
m_d\,\text{cos}\left(\frac{\pi_3}{f_\pi}-\frac{\eta_0}{f_\pi}-\,c_2^d\,\frac{\hat a}{\fpq}\right)\right]\, .
\end{align}

\item The coupling to gluons generates an effective potential that is responsible for the bulk of the mass of the $\eta'$ and can be parametrized at low energies as, 
\begin{align}
 \delta\LL_{a,\,\text{anom}}\,=-\frac{\alpha_s}{8\,\pi}\left(2\frac{\eta_0}{f_\pi}+\,c_3\,\frac{\hat a}{f_a}\right)\,G\,\tilde G\longrightarrow-\frac{1}{2}\,K\,\left(2\frac{\eta_0}{f_\pi}+\,c_3\,\frac{\hat a}{f_a}\right)^2\,.
\end{align}
\end{itemize}
Altogether, the relevant  chiral Lagrangian reads,
\begin{multline}
\delta\LL_{a}\,= \frac{1}{2}\,\dmu\hat a\,\dmup\hat a\, + 
\, \frac{1}{2}\,\dmu\pi_3\,\dmup\pi_3\, +
\, \frac{1}{2}\,\dmu\eta_0\,\dmup\eta_0\, +
\,\alpha_\pi\,\dmu\hat a\, \dmup\pi_3  \,+
\,\alpha_\eta\,\dmu\hat a\, \dmup\eta_0 \, \\
+\,B_0 f_{\pi}^2 \left[
m_u\,\text{cos}\left(\frac{\pi_3}{f_\pi}+\frac{\eta_0}{f_\pi}-\,c_2^u\,\frac{\hat a}{\fpq}\right)+
m_d\,\text{cos}\left(\frac{\pi_3}{f_\pi}-\frac{\eta_0}{f_\pi}-\,c_2^d\,\frac{\hat a}{\fpq}\right)\right] \\
\,-\,\frac{1}{2}\,K\,\left(2\frac{\eta_0}{f_\pi}+\,c_3\,\frac{\hat a}{f_a}\right)^2\, . 
 \label{Eq:chiralLag before kinetic diag}
\end{multline}
The coefficients $\alpha_\pi,\,\alpha_\eta$ parametrizing the kinetic mixing are
\begin{equation}
\alpha_\pi=\big(c_1^u-c_1^d\big)\,\frac{f_\pi}{f_{\text{PQ}}};\qquad 
\alpha_\eta=\big(c_1^u+c_1^d\big)\,\frac{f_\pi}{f_{\text{PQ}}}.
\end{equation}
In order to obtain the mass matrix and ultimately the mass eigenvalues and mixings, the kinetic terms have to be diagonalized first. This can be done by the following transformations: 
\begin{align}
\hat a\,\to  \,\frac{\hat a}{\sqrt{1-\alpha_{\pi}^2-\alpha_{\eta'}^2}}
& \simeq \hat a \, ,\nn\\
\pi_3 \,\to\, \pi_3 -\frac{\alpha_\pi \,\hat a}{\sqrt{1-\alpha_{\pi}^2-\alpha_{\eta'}^2}}
&\simeq \pi_3 - \big(c_1^u-c_1^d\big)\,\frac{f_\pi}{f_{\text{PQ}}}\, \hat a\, ,\nn\\
\eta_0\,\to\, \eta_0\,-\,\frac{\alpha_\eta \,\hat a}{\sqrt{1-\alpha_{\pi}^2-\alpha_{\eta'}^2}}
&\simeq \eta_0 - \big(c_1^u+c_1^d\big)\,\frac{f_\pi}{f_{\text{PQ}}}\, \hat a\, .\,
\label{Eq:Kinetic Diag Transf}
\end{align}
The Lagrangian Eq.~(\ref{Eq:chiralLag before kinetic diag}) becomes\footnote{The same Lagrangian can be obtained by making use of the reparametrization invariance in Eq.~(\ref{Eq:reparametrization}) and choosing $\alpha_u=c_1^u$ and $\alpha_d=c_1^d$.} ,
\begin{multline}
\delta\LL_{a}\,= \frac{1}{2}\,\dmu\hat a\,\dmup\hat a\, + 
\, \frac{1}{2}\,\dmu\pi_3\,\dmup\pi_3\, +
\, \frac{1}{2}\,\dmu\eta_0\,\dmup\eta_0\, \\
+\,B_0 f_{\pi}^2 \left[
m_u\,\text{cos}\left(\frac{\pi_3}{f_\pi}+\frac{\eta_0}{f_\pi}-\,\bar{c}_2^u\,\frac{\hat a}{\fpq}\right)+
m_d\,\text{cos}\left(\frac{\pi_3}{f_\pi}-\frac{\eta_0}{f_\pi}-\,\bar{c}_2^d\,\frac{\hat a}{\fpq}\right)\right]\\
\,-\,\frac{1}{2}\,K\,\left(2\frac{\eta_0}{f_\pi}+\,\bar{c}_3\,\,\frac{\hat a}{\fpq}\right)^2\, ,
 \label{Eq:chiralLag after kinetic diag}
\end{multline}
where $\bar{c}_2^u$, $\bar{c}_2^d$ and $\bar{c}_3$ are given by
\begin{align}
\bar{c}_2^u=& \,2\,c_1^u+\,c_2^u \, ,              \nn \\
\bar{c}_2^d=& \,2\,c_1^d+\,c_2^d \, ,             \nn \\
\bar{c}_3=& \, c_3-2\,c_1^u-\,2\,c_1^d \, . 
\end{align}
As expected, these coefficients are invariant under the reparametrization invariance in Eq.(\ref{Eq:reparametrization}). Therefore, the  squared mass matrix coming from Eq.~(\ref{Eq:chiralLag after kinetic diag}) is completely general,
\beq
\scriptstyle M^2_{\{\pi_{3},\,\eta_{0},a\}}=
\left(
\begin{array}{ccc}
\scriptstyle B_0\,(m_u+m_d)  & \scriptstyle \quad B_0\,(m_u-m_d)&
\scriptstyle -B_0 \frac{f_\pi}{\fpq} \big(m_u\,\bar{c}_2^u-m_d\,\bar{c}_2^d\big)\\ 
\scriptstyle B_0\,(m_u-m_d)  &  \scriptstyle \quad \frac{4K}{f_\pi}+B_0(m_u+m_d) &
\scriptstyle \frac{2\,\bar{c}_3 K}{f_\pi \fpq}+B_0 \frac{f_\pi}{\fpq} \big(m_u\,\bar{c}_2^u+m_d\,\bar{c}_2^d\big)\\ 
\scriptstyle -B_0 \frac{f_\pi}{\fpq} \big(m_u\,\bar{c}_2^u-m_d\,\bar{c}_2^d\big) & 
\scriptstyle \frac{2\,\bar{c}_3 K}{f_\pi \fpq}+B_0 \frac{f_\pi}{\fpq} \big(m_u\,\bar{c}_2^u+m_d\,\bar{c}_2^d\big)  &  
\scriptstyle \bar{c}_3^{\,2}\frac{K}{\fpq^2} +B_0 \frac{f_\pi^2}{\fpq^2} \big(m_u\,(\bar{c}_2^u)^{2}+m_d\,(\bar{c}_2^d)^{2}\big)
\end{array}  
\right)\,.\label{Eq:GeneralMassmatrix}
\eeq
This matrix can be diagonalized analytically in the limit $\fpq\gg B_0,\,m_{u,d},\,f_\pi$. We find that the physical axion corresponds to the combination:
\begin{align}
a & \simeq \hat{a} +\theta_{a\pi}\,\pi_3 + \theta_{a\eta'}\,\eta_0\,,
\end{align}
where all mixing angles are assumed small and 
\begin{align}
&\theta_{a\pi}\simeq -\frac{f_\pi}{2\fpq}\frac{\left(\bar{c}_3+2\bar{c}_2^d\right)m_d-\left(\bar{c}_3+2\bar{c}_2^u\right)m_u}{m_u+m_d}\,, \qquad &\theta_{a\eta'} \simeq -\frac{f_\pi}{2\fpq}\,\bar{c}_3.\qquad 
\end{align}
Note that this gives the physical axion in terms of the fields whose kinetic terms are already diagonalized. In order to express it in terms of the \emph{flavour} meson fields (as defined in Eq.~(\ref{Eq:pion and era currents})), the transformation in Eq.~(\ref{Eq:Kinetic Diag Transf}) needs to be taken into account. The final mixings read,
\begin{align}
&\theta_{a\pi}\simeq -\frac{f_\pi}{2\fpq}\left(c_2^d-c_2^u+\big(c_3+c_2^u+c_2^d\big)\frac{m_d-m_u}{m_u+m_d}\right)\,, \qquad &\theta_{a\eta'} \simeq -\frac{f_\pi}{2\fpq}\,c_3\, ,\qquad 
\end{align}
expressed in terms of the couplings of the starting Lagrangian in Eq.~(\ref{Eq:Lgeneral}). It is worth noting that the physical mixing parameters do not depend on the coefficient $c_{1}^{u,\,d}$ of the derivative operator, since it is PQ invariant and therefore it has no impact on  the masses and the mixings.

Now we are ready to study the effect of these two diagonalizations in the coupling of the axion to photons,
\beq
g_{a\gamma\gamma} = g_{a\gamma\gamma}^0 + \theta_{a\pi}\, g_{\pi \gamma\gamma}+ \theta_{a\eta'}\,g_{\eta' \gamma\gamma}\,.
\eeq
Taking into account that the coupling of the mesons to photons are given by

\begin{align}
&g_{\pi \gamma\gamma} \equiv -\frac{3\,\alpha}{\pi\,f_\pi} \big(q_u^2-q_d^2\big), 
&g_{\eta' \gamma\gamma} \equiv -\frac{3\,\alpha}{\pi\,f_\pi} \big(q_u^2+q_d^2\big),
\end{align}
where $q_u$ and $q_d$ are the electromagnetic charges of the up and down quarks, we find,

\begin{equation}
g_{a\gamma\gamma} = g_{a\gamma\gamma}^0 + \frac{\alpha}{2\pi \fpq}\left(-6\,c_2^u\, q_u^2-6\,c_2^d\, q_d^2-6\left(c_3+c_2^u+c_2^d\right)\frac{q_d^2\,m_u-q_u^2\,m_d}{m_u+m_d}\right)\, .
\end{equation}
Recalling that the coefficients $c_2^{u,\,d}$ correspond to the PQ charges,  the combinations that appear in the above equation can be identified as the electromagnetic  and QCD anomaly coefficients for the up and down quarks,
\begin{align}
&E_{u,\,d}=\sum_{\psi=u,\,d}2\, \mathcal{X}_\psi\,q_\psi^2=-6\,c_2^u\, q_u^2-6\,c_2^d\, q_d^2\, ,
&N=c_3+c_2^u+c_2^d\,.
\end{align} 
 Redefining the axion decay constant as usual $f_a= \fpq/N$, we can express
\begin{equation}
g_{a\gamma\gamma} = g_{a\gamma\gamma}^0 + \frac{\alpha}{2\pi f_a}\left(\frac{E_{u,\,d}}{N} -\frac{2}{3}\frac{m_u+4\,m_d}{m_u+m_d}\right)\,.
\end{equation}
To sum up, from the most general mass matrix we have obtained the same result of Eq.~(\ref{gagamma}) taking into account that in the $E/N$ we have to sum over all fermions transforming under the PQ symmetry, that means including the up and down quarks $E=E_{\text{heavy}}+E_{u,\,d}$,
\begin{equation}
g_{a\gamma\gamma} = \frac{\alpha}{2\pi f_a}\left(\frac{E_{\text{heavy}}}{N}+\frac{E_{u,\,d}}{N} -\frac{2}{3}\frac{m_u+4\,m_d}{m_u+m_d}\right).
\end{equation}

\newpage
\bibliographystyle{utphys.bst}
\bibliography{biblio.bib}

\end{document}